\documentclass{template}
\usepackage{amsmath}
\usepackage{cite}
\usepackage{booktabs}

\begin{document}
\pagestyle{fancy}
\pagestyle{plain}

\title{Short-Range Order and Li$_x$TM$_{4-x}$ Probability Maps for Disordered Rocksalt Cathodes} 
\maketitle

\author{Tzu-chen Liu}
\author{Steven B. Torrisi}
\author{Chris Wolverton*}

\begin{affiliations}
Tzu-chen Liu, Chris Wolverton\\
Department of Materials Science and Engineering, Northwestern University, Evanston, IL 60208, USA\\
Email Address: c-wolverton@northwestern.edu\\[1ex]

Steven B. Torrisi\\
Energy \& Materials Division, Toyota Research Institute, Los Altos, CA 94022, USA
\end{affiliations}

% Keywords: Please provide a minimum of three and a maximum of seven keywords, separated by commas


\justifying

\begin{abstract}
Short-range order (SRO) in the cation-disordered state is a controlling factor influencing the probability of finding Li$_{4}$ tetrahedron clusters in disordered rocksalt (DRX) cathode materials. 
However, the prevalent Li$_4$ probability below the random limit across reported DRX compositions has not been systematically investigated, active strategies to surpass the random limit of Li$_4$ probability are lacking, and the fundamental ordering behavior on the face-centered cubic (FCC) lattice remains insufficiently explored. 
This research quantitatively examines pair SRO parameters and Li$_x$TM$_{4-x}$ probabilities via exhaustive Monte Carlo mapping across a simplified subset of the parameter space.
The results indicate that, in the disordered state, the Li$_4$ probability is governed by the nearest neighbor (NN) pair-wise SRO parameter, and that these quantities do not necessarily represent a simple attenuation of their corresponding low-temperature long-range order, particularly for the important cases of Layered and Spinel-like orderings.
Strategies are proposed to mitigate or even reverse the lithium and transition metals mixing tendency of NN pair SRO to achieve Li$_4$ probabilities that exceed the random limit. 
This study advances the fundamental thermodynamic understanding of ordering behaviors, which can be generalized to any FCC system.

\end{abstract}

%\keywords{Suggested keywords}%Use showkeys class option if keyword
                              %display desired

%\tableofcontents

\section{\label{sec:intro}Introduction}

\begin{figure}
\centering
\includegraphics[width=0.8\columnwidth]{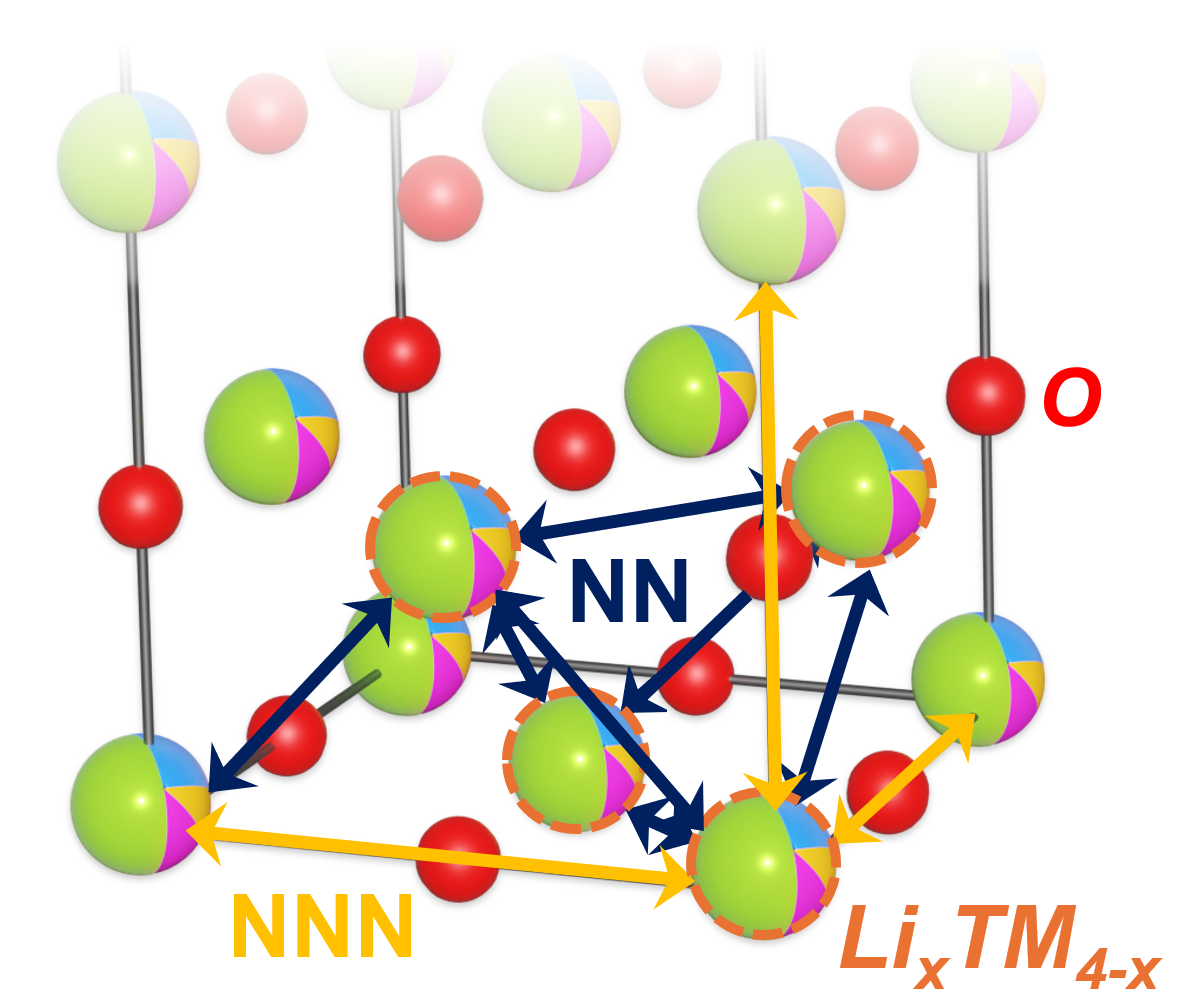}
\caption{\label{fig:str} Disordered rocksalt structure and  examples of key clusters on the FCC cation sublattice. 
The fully disordered state means that all cation FCC sites exhibit equal probability of Li$_x$TM$_{4-x}$ occupations (Li = green hemisphere, and TM = other TM mixing in our pseudobinary Li-TM scheme).  
Some nearest neighbor and next nearest neighbor pair clusters of cation sites are labeled as NN (dark blue arrows) and NNN (yellow arrows), respectively, for demonstrations. 
Four cation sites that are outlined by dotted orange lines and connected through six NN edges form the smallest tetrahedron cluster in FCC. 
In the case that all four corners of the tetrahedron are Li (termed as Li$_{4}$ in our ordering investigations), it forms the active Li diffusion ``0-TM'' channels \cite{lee2014unlocking}.  
NN and NNN pair short-range order parameters, and the Li$_x$TM$_{4-x}$ probabilities are the targeted ordering parameters, which are intercorrelated with each other due the FCC geometry constraints, as investigated in this research using Monte Carlo mapping.}
\end{figure}

The ordering behavior of lithium in Lithium Transition Metal Oxide (LiTMO$_2$) cathode materials plays a central role in key properties like rate capability, discharge capacity, and degradation mechanisms \cite{whittingham2004lithium}.
For certain emerging classes of cathodes, such as the disordered rocksalt-type (DRX) structure \cite{lee2014unlocking, clement2020cation}, order-disorder phenomena is an intrinsic part of their overall behavior, and so a thorough understanding of these phenomena is essential for material optimization.
The DRX structure has disordering of transition metals (TMs) and lithium ions on the face-centered cubic (FCC) cation sublattice, in which short‑range order (SRO) accounts for local deviations from uniformly random configurations.
Of particular importance for Li ordering behavior in DRX is the probability of forming Li$_{4}$ \cite{ji2019hidden}, the smallest four-body cluster (tetrahedron) with Li occupying all cation sites \cite{lee2014unlocking} (see local motifs labeled ``0-TM''; further explained in Figure \ref{fig:str}).
This probability is crucial for any candidate DRX chemistry under study, as it promotes the formation of percolating channels \cite{urban2014configurational} for facile Li diffusion and makes a decisive difference in electrochemical cycling performance.

However, there are practical challenges to understanding and controlling Li ordering, for both computations or experiments.
Firstly, the ordering behavior of Li varies with the other cation species that are present, meaning individual systems require their own consideration.
Nevertheless, even when studying one particular chemistry, the common modeling approach of the standard cluster expansion (CE) method \cite{sanchez1984generalized,connolly1983density} relies on time-consuming and data-intensive procedures. 
Second, previous DRX studies have overwhelmingly reported an unfavorable SRO tendency compared to the random limit and considered SRO generally detrimental to Li$_4$ probability \cite{clement2020cation, ji2019hidden,  kitamura2020local, ouyang2020effect, lun2020design,  lun2021cation, cai2022thermodynamically, squires2023understanding, szymanski2023modeling, li2023atomic, chen2024exploring, zhang2024screening}, except in cases of significant fluorination (around 30 \%) \cite{ouyang2020effect, lun2020design}. 
Li$_4$ probabilities below the random limit are commonly observed alongside Li$_2$TM$_2$ probabilities above the random limit (the tetrahedron with maximal Li–TM mixing). 
This phenomenon has been referred to as $\gamma$-LiFeO$_2$–type SRO \cite{lun2021cation, chen2024exploring}, with $\gamma$-LiFeO$_2$ being a structure composed exclusively of Li$_2$TM$_2$ tetrahedra (see Table \ref{tab:Corr_table}).
Without successful examples of favorable SRO to promote Li$_4$-rich environments in pure oxides, no useful strategy has been derived for designing cathode chemistries that effectively surpasses the random limit of Li$_4$ probability; the only seemingly practical strategy is the suppression of unfavorable SRO to approximate the random limit as closely as possible \cite{lun2021cation, cai2022thermodynamically, squires2023understanding}.
Here, we demonstrate how these challenges may be clarified and addressed by a simplified but effective CE framework with Monte Carlo mapping that disentangles the key contributions to SRO behavior on the FCC cation lattice in LiTMO$_2$. 
The framework also helps to elucidate chemical design considerations for DRX cathodes to surpass the random limit of Li$_4$ probability, which is unlikely to be achieved by brute-force explorations.

%%%%%%%%%%%%%%%%%%%%%%%%%%%%%%%%%%%%%%%%%%%%%%%%%%%%%%%%%%%%%%%%%%%%%%%%%%%%%%%%%%%%%%%%%%%%%%%%%%%%%%%%%%%%%%%%
Before introducing our quantitative SRO analysis, we point out two distinctions from conventional SRO studies: 
(i) the ordering objective centers on a four-body cluster, a motif less explored than pair SRO in classical ordering research, and 
(ii) the cluster’s probability deviating from the random limit only in one direction (in this case, Li–TM mixing) across compositions is somewhat uncommon. 
The SRO is traditionally parametrized for pair clusters in binary alloys \cite{cowley1950approximate, cowley1960short} by $\alpha_{(n)}^{AB}(x,T)$: this two-body quantity reflects the statistical tendency of microstates exhibiting either clustering (more A-A, B-B pairs) or ordering/mixing (more A-B pairs), quantified as fluctuations above or below the random distribution \cite{wolverton2000short}:

\begin{equation}
    \label{SRO_p}
    \alpha_{(n)}^{AB}(x,T) = 1 - \frac{P_{(n)}^{AB}(x,T)}{x}
\end{equation}
where $P_{(n)}^{AB}(x,T)$ denotes the conditional probability of finding a $B$ atom in the $n^{th}$ shell given that an $A$ atom is at the origin, as a function of composition $x$ and temperature $T$. 
At the random limit, $ P_{(n)}^{AB}(x,T) = x$ and so $\alpha_{(n)}^{AB}(x,T) =0$; as such this quantity can be positive/negative to indicate clustering/mixing respectively. Note that this parameter remains well defined even in the presence of long‑range order (LRO). 
While there are no commonly-used SRO parameters for higher-order clusters, conceptually the SRO can still be used to describe phenomena regarding the probability of finding any specific multi-site cluster configurations relative to the random distribution. 
Moreover, the tetrahedron cluster probability is analytically linked to its subcluster SRO parameters \cite{ceder1991alloy,de1992cluster}, as discussed in Figure \ref{fig:str} and Section \ref{sec:formalism}.  

In principle, the observation that four-body SRO almost always acts as a detrimental factor (less Li${_4}$ than the random limit) is puzzling from a conventional SRO analysis perspective, as fluctuations should theoretically also include the chance of being favorable (promoting more Li${_4}$ than the random limit).
Across reported compositions, a commonly observed negative bias of Li${_4}$ probability below the random limit and prevalent Li-TM mixing SRO is often explained by electrostatic considerations. 
These arguments are intuitively reasonable: Li$^{+}$ and TM$^{3+}$ ions tend to mix to maintain local charge neutrality \cite{ji2019hidden,jones2019short}, whereas elastic or strain interactions \cite{clement2020cation, ji2019hidden, lun2021cation}, due to the ionic size mismatch between Li and the TMs, acts as a counterpart to explain why some systems outperform others in Li$_4$ clustering, though all noticeably worse than the random limit.
More interestingly, there is an intuition that SRO in the high-temperature disordered state represents ``remnants/precursors'' of the corresponding low-temperature ground-state LRO \cite{ji2019hidden, wolverton2000short}.
The common symmetry alignment between SRO and LRO is supported by the mean-field theory \cite{krivoglaz1969theory, clapp1966correlation}, with rare exceptions investigated in dedicated studies\cite{wolverton1994long, lu1994unequal, wolverton1995first, wolverton2000short}. 
Since the random state is the infinitely high-temperature limit ($T=\infty$), the SRO in the finite-temperature disordered state ($\infty>T>T_c$) is reasonably hypothesized as the intermediate state by assuming monotonic changes between the ground state and the random limit. 
An extended hypothesis for the mathematical representation of this intuition is introduced in the CE formalism section.
If electrostatically favored Li$_2$TM$_2$ and Li-TM mixing were dominant in LiTMO$_2$ as hinted by the SRO observations, one might expect the frequent appearance of the electrostatic ground-state LRO ($\gamma$-LiFeO$_2$-type) \cite{magri1990ground, wolverton1995short, wu1998size}; in reality, however, this LRO type is rarely reported as 0 K stable state in nature \cite{hewston1987survey}.
Lun~\textit{et al.}\cite{lun2021cation} have tentatively attributed this contradiction to the long-range nature of elastic contributions, mirroring the argument in the alloy community for explaining conflicts between LRO and SRO caused by significant strain in coherent phase separation \cite{wolverton2000short, lu1994unequal,  smith2024competition}; 
however, direct evidence supporting this explanation for the distinct trend of LRO vs SRO in LiTMO$_2$ is still lacking, even at a qualitative level.
Here, we demonstrate that the above puzzle can be quantitatively addressed by understanding the fundamental nature of pair SRO parameters and the analytic Li$_4$ probability on the FCC lattice, irrespective of the interactions’ physical origin.

%%%%%%%%%%%%%%%%%%%%%%%%%%%%%%%%%%%%%%%%%%%%%%%%%%%%%%%%%%%%%%%%%%%%%%%%%%%%%%%%%%%%%%%%%%%%%%%%%%%%%%%%%%%%%%%%
In this paper, we present a simplified CE formalism coupled with exhaustive Monte Carlo (MC) simulations, enabling comprehensive mapping of the first two (NN and NNN) pair SRO parameters and the smallest four-body cluster probabilities (denoted as Li$_x$TM$_{4-x}$ probabilities hereafter). 
We aim to investigate the general tendency of these quantities and their interconnections as a function of their effective cluster interactions (ECIs). 
Based on the literature and a series of exemplary CE models we analyzed, we find the evident role of short-range clusters in projecting ordering energetics on the cluster basis. 
A previous DRX study \cite{liu2025tailored} produced a dataset containing 6,182 compositions and 24,728 supercells of LiTMO$_2$ with two and three mixed TMs from the Open Quantum Materials Database (OQMD) \cite{saal2013materials, kirklin2015open}.
Here, we leverage this large dataset to statistically estimate the distribution of pseudobinary Li-TM ECIs, in order to determine the important ECIs' range to focus on and to build the connection between ECIs and LiTMO$_2$ energies. 

Under the approximation of pseudobinary ECIs on short-range clusters, we show that most LiTMO$_2$ systems that form any single ordered phase as their low-temperature ground states (either Spinel-like, Layered, $\gamma$-LiFeO$_2$, or CuAu-type ordering), their high-temperature disordered states are destined to exhibit Li${_4}$ probability noticeably lower than the random limit, rationalizing the Li${_4}$ deficiency frequently reported in LiTMO$_2$. 
Such unfavorable ordering behavior arises from prevalent positive NN pair interactions and mixing‑type NN pair SRO, which also lead to Li$_2$TM$_2$ probabilities above the random limit. 
We analytically investigate the Li$_4$ probability as a function of its subcluster correlations. 
Our analysis demonstrates that, the Li$_4$ and other tetrahedron cluster probabilities at $\infty>T>T_c$ for a given chemistry cannot always be inferred from intermediate values between those in its $T=0$ K ground state and $T=\infty$ random state, particularly for the important cases of Layered and Spinel-like orderings.
Moreover, by comprehensively examining the temperature‐dependent behavior of first‑two pair SRO, we reveal that, even for the NN pair SRO, the common intuition of viewing it as the ``precursor'' of LRO is an oversimplification under FCC frustration effects.
Finally, we discuss options to improve the Li${_4}$ probability in thermodynamically equilibrated states. 
We note that, although this work is motivated by understanding the nature of pair SRO and Li$_x$TM$_{4-x}$ probabilities on the FCC cation sublattice of LiTMO$_2$ cathodes, ordering on the FCC lattice represents a well-established research area extensively studied 
\cite{cowley1950approximate, cowley1960short, wolverton1994long, wolverton1995short, wolverton1995first, wolverton2000short, burvsik1999study, rempel1990short, sanchez1981theoretical, ducastelle1991order, clapp1966correlation, gahn1975short, sanchez1980ordering, phani1980monte, sanchez1982comparison, rempel1985relation, rao2022analytical}. 
The main output, consisting of the SRO and tetrahedron cluster probability map from this work, is a natural extension of those earlier studies and can be generalized to any FCC ordering research regardless of materials chemistry.  

\section{\label{sec:formalism}Cluster Expansion formalism}
To study the ordering phenomenon on a fixed lattice, the CE formalism, a generalized Ising model, is an elegant representation of configurational degrees of freedom in terms of the orthogonal cluster basis and can be employed to represent the energetics as function of orderings \cite{sanchez1984generalized,connolly1983density,van2002automating, van2009multicomponent}: 

\begin{equation}
E(\sigma)=\sum_{\alpha} m_{\alpha} J_{\alpha} \langle \Gamma_{\alpha}(\sigma)\rangle,\
    \langle \Gamma_{\alpha}(\sigma)\rangle= \left\langle
  \prod_{i \in \alpha}
  \sigma_i
\right\rangle
    \label{CE}
\end{equation}
where $E(\sigma)$ denotes the energy of ordering configuration $\sigma$; $\alpha$ represents each symmetrically distinct cluster, labeled by the numerical pair $$(o, n)$$ in the following context, where $o$ is the order of the cluster and $n$ is its ranking within that order, arranged from short to long; $m_\alpha$ gives the symmetrical multiplicities for each cluster $\alpha$; 
$J_\alpha$ is the energy expansion coefficient, commonly termed effective cluster interactions (ECIs); 
and the $\langle \Gamma_{\alpha}(\sigma)\rangle$ denotes the cluster's correlation function, computed as the product of the occupation variables $\sigma_i$ = $\pm 1$ (depending on the element type on site $i$; Li is always assigned -1 in this work) for all sites in $\alpha$, and averaged over all clusters equivalent to $\alpha$. $\langle\Gamma_{0}\rangle$ for the null cluster is always 1. 

The Hamiltonian from Equation~\ref{CE}, with ECIs input manually adjusted to scan the entire parameter space investigated in this work, serves as the input for the following MC simulations to construct the SRO and Li$_x$TM$_{4-x}$ maps. 
Under this CE formalism, correlations of previously studied ground-state LRO of the Ising model\cite{sanchez1981theoretical, ducastelle1991order}, compared to the $T=\infty$ random state and phase-separated mixture of Li oxide and TM oxide in the rocksalt-type structure (denoted as LiO+TMO from here on), are shown in Table~\ref{tab:Corr_table}. 
We note that, although common Layered and Spinel-like orderings may appear segregated visually \cite{ ji2019hidden, urban2014configurational}, their first two pair correlations correspond to random ($\langle\Gamma_{2,1}\rangle =0$) and fully mixing ($\langle\Gamma_{2,2}\rangle = -1$) on NN and NNN sites, respectively; hence, under conventional ordering analysis, terming these structures segregating may be ambiguous. 
The intuition discussed previously---that SRO represents remnants of LRO---can be generalized and quantified as the following question and hypothesis:
Do these ordering quantities for $\infty>T>0$ K approach the random limit monotonically as temperature increases and entropic contributions become more significant?
Or, under a weaker hypothesis, for systems exhibiting any ground-state LRO, do their disordered states above $T_c$ exhibit intermediate values of correlations and cluster probabilities between the random limit and their $T=0$ K ground states on FCC lattice? 
Specifically, for systems whose $T=0$ K ground states are Layered and Spinel-like structures, $\langle\Gamma_{2,1}\rangle= 0$ identical to the $T=\infty$ random-limit value, it is interesting to learn whether $\langle\Gamma_{2,1}\rangle$ remains zero at $\infty>T>T_c$. 
Our answers to these questions helps elucidate the discrepancies in ordering observations between ordered and disordered states. 
% The answer is not always correct for tetrahedron cluster probabilities, and not even for the simple first-neighbor pair SRO $\alpha_{2,1}$ in chemistries with Layered and Spinel-like ground states.

\begin{figure}
\includegraphics[width=\columnwidth]{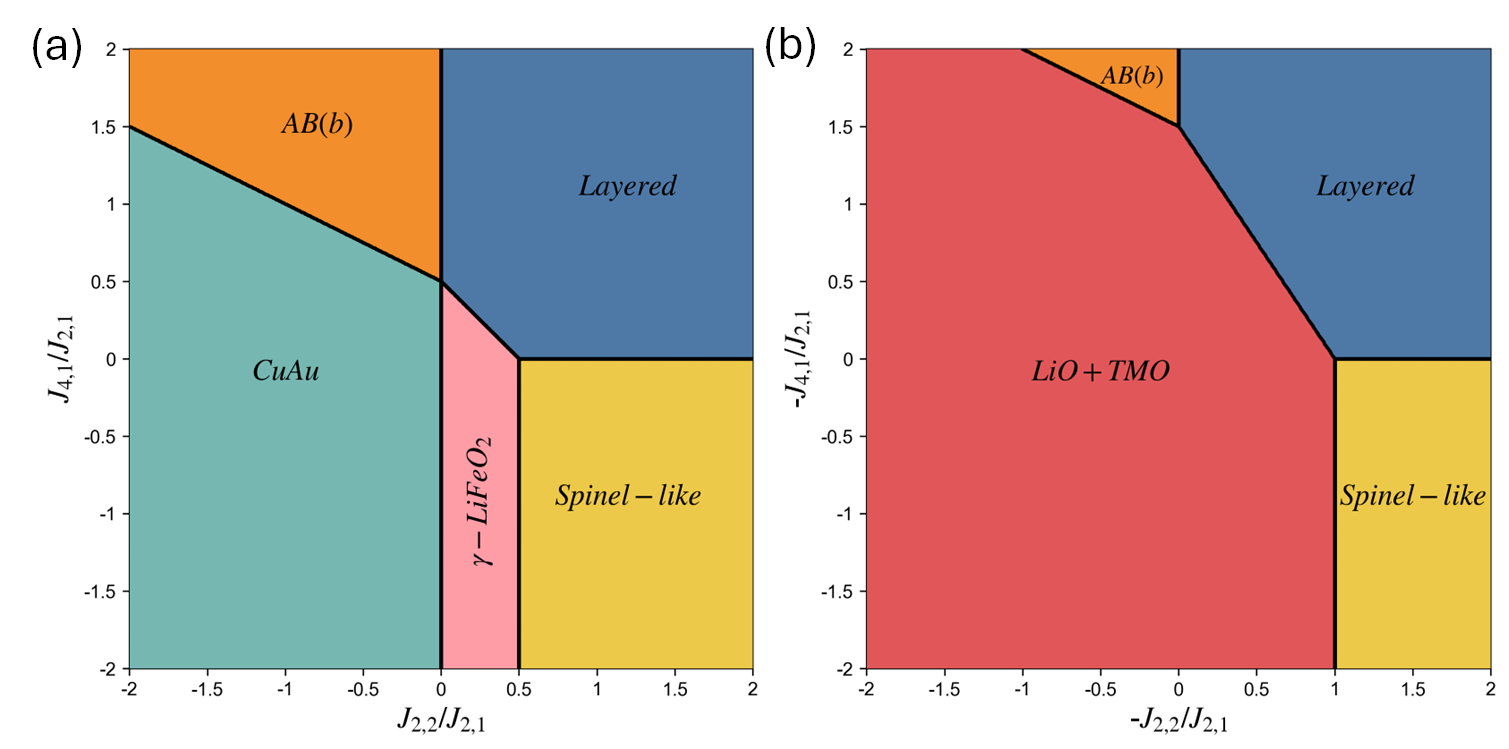}
\caption{\label{fig:GS} Ground-state diagram at Li:TM = 1:1 with (a) $J_{2,1}$ $>$ 0 (b) $J_{2,1}$ $<$ 0. Note that not all reported ground-state LRO of the Ising model appear in these diagrams, as stabilizing them requires six-body interactions \cite{sanchez1981theoretical}.}
\end{figure}

The corresponding ground-state diagrams are plotted as functions of the ECI ratios, $J_{2,2}$/$J_{2,1}$ and $J_{4,1}$/$J_{2,1}$, which are the natural coordinates since multiplying all interactions by a positive constant factor yields the same ground state determined in Figure \ref{fig:GS}. 
These ground-state diagrams serve as the inspiration for our MC mapping of pair SRO and tetrahedron cluster probabilities in terms of ECIs, enabling us to understand their connections to the corresponding ground-state configurations. 

\begin{table}
  \centering
  \caption{Correlations and tetrahedron cluster probabilities $P(\mathrm{Li}_x\mathrm{TM}_{4-x})$ for ground-state LRO of the Ising model (with NN and NNN pair, tetrahedron and octahedron many-body interactions) at Li:TM = 1:1 \cite{sanchez1981theoretical, ducastelle1991order}, compared to the random state and phase-separated mixture of LiO+TMO.}
  \label{tab:Corr_table}
  \resizebox{\linewidth}{!}{
  \begin{tabular}[h]{@{}ccccccccccc@{}}
    \hline
     & \shortstack[c]{$T=\infty$\\Random}  & Layered & Spinel-like & $\gamma$-LiFeO$_2$ & CuAu & AB(a) & AB(b) & AB(d) & AB(e) & \shortstack[c]{Phase-separated\\mixture LiO+TMO}\\
    \hline
$\langle\Gamma_{0}\rangle$ & 1       & 1   & 1    & 1     & 1    & 1    & 1    & 1    & 1    & 1\\
$\langle\Gamma_{1,1}\rangle$ & 0     & 0   & 0    & 0     & 0    & 0    & 0    & 0    & 0    & 0  \\
$\langle\Gamma_{2,1}\rangle$ & 0     & 0   & 0    & -1/3  & -1/3 & 1/3  & 0    & 0    & 0    & 1  \\
$\langle\Gamma_{2,2}\rangle$ & 0     & -1  & -1   & 1/3   & 1    & 1/3  & 1/3  & -1/3 & -1/3 & 1\\
$\langle\Gamma_{3,1}\rangle$ & 0     & 0   & 0    & 0     & 0    & 0    & 0    & 0    & 0    & 0 \\
$\langle\Gamma_{4,1}\rangle$ & 0     & -1  & 1    & 1     & 1    & 1    & -1   & 1    & -1   & 1 \\
    \hline
$P(\mathrm{Li_4})$           & 1/16  & 0   & 1/8  & 0     & 0    & 1/4  & 0    & 1/8  & 0   & 1/2   \\
$P(\mathrm{Li_3TM})$         & 4/16  & 1/2 & 0    & 0     & 0    & 0    & 1/2  & 0    & 1/2 & 0 \\
$P(\mathrm{Li_2TM_2})$       & 6/16  & 0   & 6/8  & 1     & 1    & 1/2  & 0    & 6/8  & 0   & 0  \\
    \hline
  \end{tabular}}
\end{table}

Within CE, the pair SRO parameter formula given in Equation \ref{SRO_p} (with the $n^{th}$ shell notation translated to $(o,n)$ notation in this work, and with the AB and $(x,T)$ expressions simplified) can be written as \cite{ducastelle1991order,wolverton1995short}:

\begin{equation}
    \label{SRO_c}
    \alpha_{2,n} = \frac{\langle\Gamma_{2,n}\rangle-\langle\Gamma_{1,1}\rangle^2}{1-\langle\Gamma_{1,1}\rangle^2}
\end{equation}
where $\langle\Gamma_{2,n}\rangle$ is the $n^{th}$ pair correlation, and $\langle\Gamma_{1,1}\rangle$ is the only point correlation in a binary system. 
At Li:TM = 1:1, $\langle\Gamma_{1,1}\rangle=0$, and the equation can be simplified as $\alpha_{2,n}$ = $\langle\Gamma_{2,n}\rangle$.
The Li$_x$TM$_{4-x}$ probabilities can be analytically expressed in terms of the correlations of all subclusters in the tetrahedron, as shown in the following configuration matrix (C-matrix) \cite{ceder1991alloy,de1992cluster}:
\begin{equation}
\label{C_matrix}
\begin{pmatrix}
P(\mathrm{TM_4}) \\
P(\mathrm{Li_1TM_3}) \\
P(\mathrm{Li_2TM_2}) \\
P(\mathrm{Li_3TM}) \\
P(\mathrm{Li_4})
\end{pmatrix}
=
\frac{1}{2^4}\,
\begin{pmatrix}
1 & 4 & 6 & 4 & 1 \\
4 & 8 & 0 & -8 & -4 \\
6 & 0 & -12 & 0 & 6 \\
4 & -8 & 0 & 8 & -4 \\
1 & -4 & 6 & -4 & 1 
\end{pmatrix}
\begin{pmatrix}
1 \\ \langle\Gamma_{1,1}\rangle \\ \langle\Gamma_{2,1}\rangle \\ \langle\Gamma_{3,1}\rangle \\ \langle\Gamma_{4,1}\rangle
\end{pmatrix}
\end{equation}
with its derivation outlined in the Supporting Information. 
We now understand that our focus metric, the Li$_4$ probability P(Li$_4$) in LiTMO$_2$, can be calculated analytically from:
\begin{equation}
\label{eq:Li4}
P(\mathrm{Li_4}) = \frac{1}{16}\Bigl[ 1
         - 4\,\langle\Gamma_{1,1}\rangle 
         + 6\,\langle\Gamma_{2,1}\rangle 
         - 4\,\langle\Gamma_{3,1}\rangle 
         + \langle\Gamma_{4,1}\rangle \Bigr],
\end{equation}
where $\langle\Gamma_{1,1}\rangle=0$ again and $\langle\Gamma_{3,1}\rangle$ is near zero in our simplified CE (discussed in Section \ref{sec: CE}), only the terms $\langle\Gamma_{2,1}\rangle$ = $\alpha_{2,1}$ and $\langle\Gamma_{4,1}\rangle$ remain as the primary factors impacting the Li$_4$ probability.
This equation will be used repeatedly in our analysis of MC maps in Section \ref{sec: maps}.

\section{\label{sec:method}Computational Methods}
\subsection{Density Functional Theory calculations and single TM CE models}
Energies of the training configurations for single TM CE models were obtained from density functional theory (DFT) calculations performed using the Vienna Ab initio Simulation Package (VASP) \cite{kresse1993ab, kresse1996efficiency, kresse1996efficient} with Projector Augmented Wave (PAW) pseudopotentials \cite{blochl1994projector, kresse1999ultrasoft} and the PBE exchange-correlation functional \cite{perdew1996generalized}. Plane-wave basis set cutoff energies of 520 eV for structural relaxations and 680 eV for static calculations were employed. Gamma-centered k-point grids were generated using a KSPACING value of 0.15 $\text{\AA}^{-1}$. Onsite Hubbard U corrections were applied to the d-orbitals using the Dudarev simplified rotationally invariant approach \cite{dudarev1998electron}, with the U values for each transition metal taken from the fitting of oxidation energy by Wang et al. \cite{wang2006oxidation}. The output from these calculations is used to build exemplary CE models shown in Section \ref{sec: CE} using the integrated cluster expansion toolkit (ICET) \cite{aangqvist2019icet} package.

\subsection{Monte Carlo Simulations for SRO and Li$_x$TM$_{4-x}$ Probability Maps}
All canonical MC simulations were performed using the emc2 function of the Alloy Theoretic Automated Toolkit (ATAT) \cite{van2002self}. Ensemble averaged quantities were computed using a simulation cell of 32$^3$ sites, with 10,000 MC flips per site for both equilibration and averaging. 
Cell-size convergence was tested using cells with 24$^3$, 32$^3$, and 40$^3$ sites, which yielded consistent results. 
The null and point cluster interactions are set to 0, as both terms do not contribute energy difference between microstates in canonical MC simulations. 
Without loss of generality, we set $J_{2,1}$ = 0.1 eV/site for most calculations and scan through the parameter space of $J_{2,2}$/$J_{2,1}$ and $J_{4,1}$/$J_{2,1}$, except in Figure \ref{fig:theta}, where J = 0.1 eV/site, $J_{2,1}$ = Jcos($\theta$), $J_{2,2}$ = Jsin($\theta$) (an additional scheme for a comprehensive scan of all possible $J_{2,2}$/$J_{2,1}$ ratios, including both sign of $J_{2,1}$).
By considering the equivalence of MC simulations with $k_BT/E$, all variables are normalized to a dimensionless framework: $k_BT/|J_{2,1}|$ for the temperature scale, and $J_{2,2}$/$J_{2,1}$ and $J_{4,1}$/$J_{2,1}$ as natural coordinates for our maps of averaged quantities as a function of ECIs ($k_BT/$J and $\theta$ for Figure \ref{fig:theta}). 
The results can be easily transformed to the corresponding scale for a given $J_{2,1}$ value from real LiTMO$_2$ systems. Simulations are conducted in a simulated annealing style, first in the standard run with a step of $k_B\Delta T/(|J_{2,1}|$ or J)  = 0.0431 (50 K when $|J_{2,1}|$ or J = 0.1 eV/site). 
Both temperature increasing and decreasing runs were carried out to detect potential hysteresis. Targeted SRO and Li$_x$TM$_{4-x}$ quantities at $T/T_c$ = 1.1 (representing the just-disordered state) in Section \ref{sec: maps} were obtained in the additional fine run that employed a temperature grid with steps of 0.01 $T_c$ between 1.2 and 0.8 $T_c$ (from standard run), and the $T_c$, 1.1 $T_c$, and final plotted quantities were re-determined from this fine run. 
For cases where higher variations of averaged quantities were observed in the standard run, we increased both the equilibration and averaging processes to 20,000 MC flips per site to further reduce the standard error of the observables and employed the 40$^3$-site cell to enhance the confidence in the fine run.
$T_c$ for both standard and fine runs are determined by the temperature at which heat capacity $V\!ar(E)/k_BT^2$ is maximized.

\section{\label{sec: CE} Energetic Contributions of Short-Range Clusters in LiTMO$_2$}

\begin{figure} [h]
\includegraphics[width=\columnwidth]{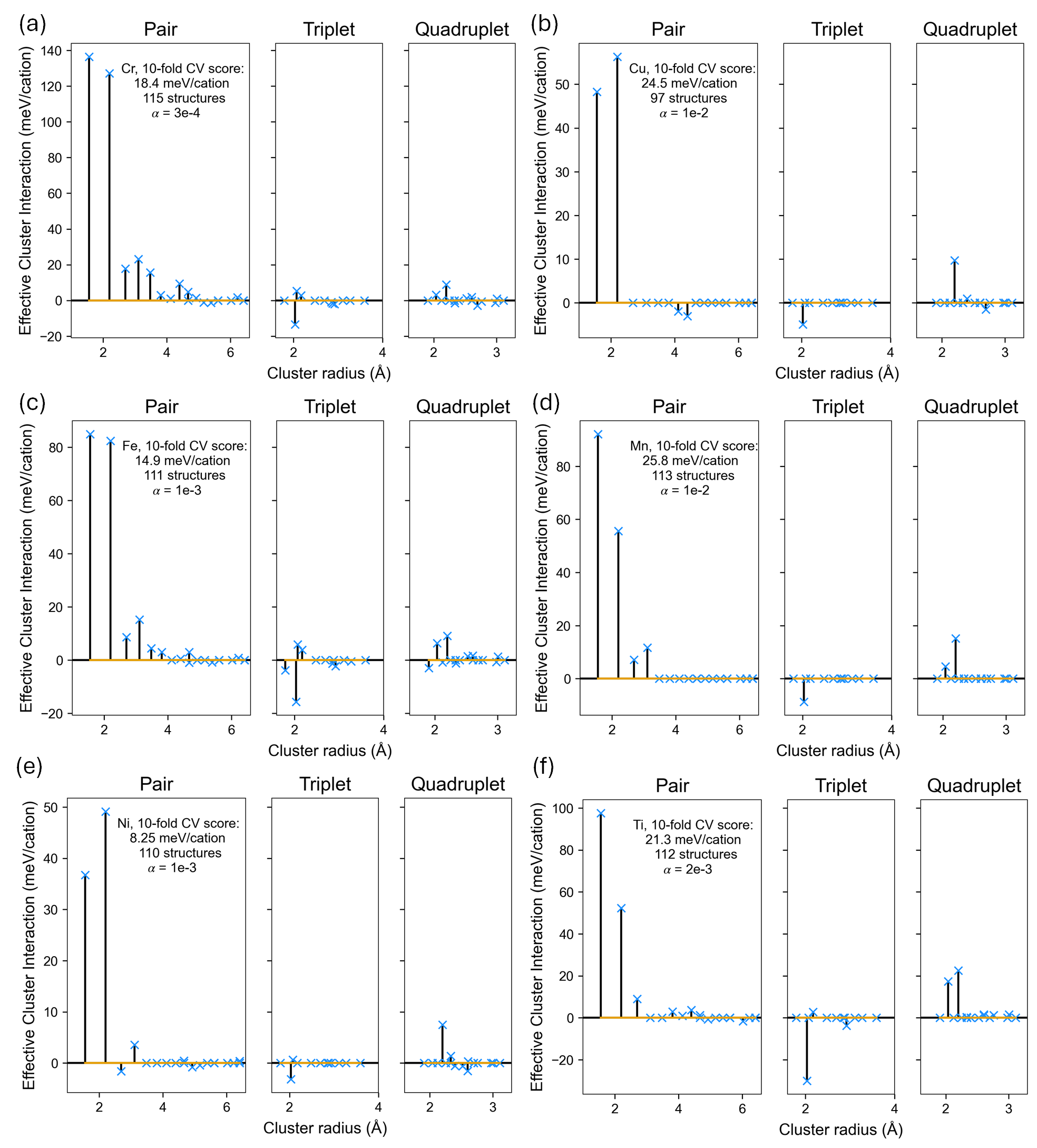}
\caption{\label{fig:ECIs} ECI plots for single TM CE corresponding to (a) Cr, (b) Cu, (c) Fe, (d) Mn, (e) Ni, and (f) Ti demonstrate the major energetic contributions from short-range clusters, along with the corresponding 10-fold cross-validation (CV) scores, numbers of training structures, and the penalty term $\alpha$ employed in Adaptive-LASSO for ECIs fitting, as determined by hyperparameter scanning for the lowest CV score.}
\end{figure}

To bridge the gap between MC maps and LiTMO$_2$, we first identify clusters with significant energetic contributions in LiTMO$_2$, while ensuring that the model remains sufficiently simple for exhaustive parameter scans in our following MC mappings. 
We note that the following standard cluster expansions are used solely to identify the key clusters for our approximations; although well-converged, they are not intended for pursuing highly refined models, nor are they employed in our MC mapping scheme.
As shown in Figure \ref{fig:ECIs}, for LiTMO$_2$ (TM = Cr, Cu, Fe, Mn, Ni and Ti), similar to previously reported Co and V cases \cite{wolverton1998cation, guo2023intercalation}, the NN and NNN pair ECIs are commonly the most dominant terms in the expansion. 
Although non-negligible contributions from longer-range clusters may exist, they are likely to have less impact on our targeted ordering parameters than the first two pairs, as discussed in the Supporting Information.
Therefore, we consider the first two pair ECIs as the major variables in the following MC mappings. 

While the first four-body ECI might not be prominent in many CE, the special role of Li$_4$ tetrahedron cluster as a key target in DRX cathodes motivates the inclusion of energetic changes directly associated with this cluster in our maps.
Moreover, the first tetrahedron cluster ($o,n$) = (4,1) is the smallest term in cluster basis to distinguish the Layered and Spinel-like orderings (which exhibit same correlations for all pairs and three-body clusters) and the corresponding four-body ECI breaks the degeneracy of these two common orderings, as shown in Figure \ref{fig:GS}. 
These key clusters are then sufficient to illustrate the stabilization of common ground-state LRO observed in LiTMO$_2$ cathode materials (i.e., Layered, Spinel-like, and $\gamma$-LiFeO$_2$ type), and we aim to thoroughly investigate how the same set of short-range interactions leads to SRO at $T>T_c$. 
We note that, without three-body cluster interactions, our model lacks antisymmetric features between Li-rich and TM-rich local environments; accordingly, $\langle\Gamma_{3,1}\rangle$ is effectively zero for all converged simulations.  
As a result, only $\langle\Gamma_{2,1}\rangle$ and $\langle\Gamma_{4,1}\rangle$ in Equation \ref{eq:Li4} determine the final Li$_4$ probability we analyzed in this work.

After selecting the first two pair and the first four-body ECIs as input variables for complete MC scanning, we estimate the common range of ECIs in LiTMO$_2$ compounds by leveraging the LiTMO$_2$ dataset with 6182 compositions in OQMD to project their ordering energetics onto these key cluster interactions and analyze their distributions. 
The corresponding simplified CE Hamiltonian used as the MC input is then given by:

 \begin{equation}\label{E_p2p1}
\begin{split}
  E &= J_{0}
      + J_{1,1}\langle\Gamma_{1,1}\rangle
      + m_{2,1}J_{2,1}\langle\Gamma_{2,1}\rangle
      + m_{2,2}J_{2,2}\langle\Gamma_{2,2}\rangle
      + m_{4,1}J_{4,1}\langle\Gamma_{4,1}\rangle \\[6pt]
    &= J_{0}
      + 6\,J_{2,1}\langle\Gamma_{2,1}\rangle
      + 3\,J_{2,2}\langle\Gamma_{2,2}\rangle
      + 2\,J_{4,1}\langle\Gamma_{4,1}\rangle
\end{split}
\end{equation}

From Table \ref{tab:Corr_table}, we can use the simplified expansion to express the energy of ordered structures as follows:
\begin{equation}
\label{DRX_projection}
\begin{pmatrix}
E_{\mathrm{Random}}\\
E_{\mathrm{Layered}}\\
E_{\mathrm{Spinel\text{-}like}}\\
E_{\gamma\text{-LiFeO}_2}
\end{pmatrix}
=
\begin{pmatrix}
1 & 0 & 0 & 0  \\
1 & 0 & -1 & -1 \\
1 & 0 & -1 & 1 \\
1 & -1/3 & 1/3 & 1 
\end{pmatrix}
\begin{pmatrix}
J_0 \\ 6J_{2,1} \\ 3J_{2,2}\\2J_{4,1}
\end{pmatrix}
\end{equation}
Inversely,
\begin{equation}
\label{DRX_inv}
\begin{pmatrix}
J_0 \\ 6J_{2,1} \\ 3J_{2,2}\\2J_{4,1}
\end{pmatrix}
=
\begin{pmatrix}
1 & 0 & 0 & 0  \\
4 & -2 & 1 & -3 \\
1 & -1/2 & -1/2 & 0 \\
0 & -1/2 & 1/2 & 0 
\end{pmatrix}
\begin{pmatrix}
E_{\mathrm{Random}}\\
E_{\mathrm{Layered}}\\
E_{\mathrm{Spinel\text{-}like}}\\
E_{\gamma\text{-LiFeO}_2}
\end{pmatrix}
\end{equation}
where energies for Layered, Spinel-like, and $\gamma$-LiFeO$_2$ structures are extracted from the OQMD database. 
We again note that these projected ECIs are neither intended nor sufficiently precise to replace standard CE as accurate models. 
Instead, we utilize their distributions to guide the ECI's range of exhaustive scans and link ECIs to the real energies of thousands of LiTMO$_2$, thereby offering insight into their SRO in the comprehensive chemical space that is prohibitively expensive to study thoroughly with standard CE.
The effectiveness and the approximate nature of the projected ECIs are further discussed in the Supporting Information.

\begin{figure} [h]
\includegraphics[width=\columnwidth]{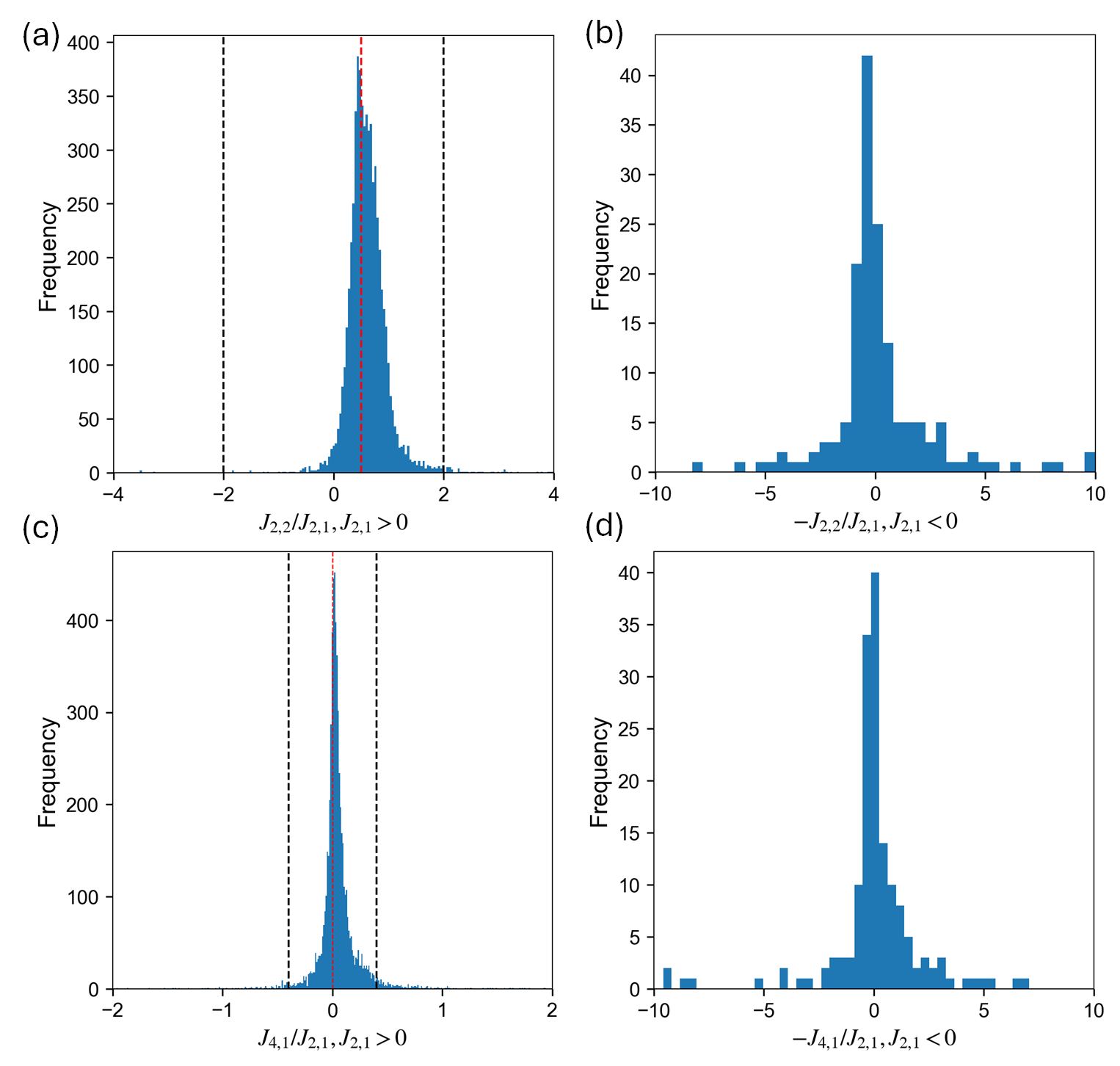}
\caption{\label{fig:OQMD_Hist} Histogram of the projected $J_{2,2}$/$J_{2,1}$ and $J_{4,1}$/$J_{2,1}$  for (a)(c) 6019 compositions with $J_{2,1}$ $>$ 0 (b)(d) 163 compositions with $J_{2,1}$ $<$ 0. The red lines in $J_{2,1} > 0$ plots represents $J_{2,2}$/$J_{2,1}$ = 0.5 and 0 for $J_{2,2}$/$J_{2,1}$ and $J_{4,1}$/$J_{2,1}$, respectively, which is around the peak of distributions. The black dotted lines indicate the region of $J_{2,2}$/$J_{2,1}$ values between $\pm$2 and $J_{4,1}$/$J_{2,1}$ values between $\pm$0.4 in the positive $J_{2,1}$ plots, covering 5937 and 5701 compositions, respectively. These intervals encompass the majority of the enumerated dataset and will be the focus of the subsequent MC mappings.}
\end{figure}

The resulting distributions of $J_{2,2}/J_{2,1}$ and $J_{4,1}/J_{2,1}$ derived from projections of 6,182 compositions are shown in Figure \ref{fig:OQMD_Hist}. 
Nearly 97\% of enumerated compositions exhibit projected $J_{2,1}>0$, promoting Li-TM mixing ($\langle\Gamma_{2,1}\rangle < 0$). 
For compositions with $J_{2,1}>0$, the $J_{2,2}/J_{2,1}$ distribution peaks near 0.5, coinciding with the highly degenerate point in Figure \ref{fig:GS}a and indicating that their energies in the $\gamma$‑LiFeO$_2$, Layered, and Spinel‑like structures are comparable and, as shown by the second and third rows of Equation \ref{DRX_inv}, lie below $E_{\mathrm{Random}}$.
Nearly 99\% of entries with $J_{2,1}>0$ fall within $J_{2,2}/J_{2,1}\in[-2,2]$, corresponding to the x‑axis of Figure \ref{fig:GS}a, and will serve as our primary research target in the next section. 

Next, the $J_{4,1}/J_{2,1}$ distribution ($J_{2,1}>0$) of entries exhibits a particularly narrow distribution that centers around 0, reflecting the previously reported trend \cite{liu2025tailored} that Layered and Spinel‑like energies are nearly identical in most compositions. 
Nevertheless, it remains interesting to examine the effect of this four‑body ECI term on the tetrahedron cluster probability, as previously discussed, albeit at a smaller magnitude than the $J_{2,2}$ term.
Finally, we briefly discuss the less common subset with $J_{2,1}<0$ (3\%, 163 compositions). 
Their $-J_{2,2}/J_{2,1}$ and $-J_{4,1}/J_{2,1}$ distributions also peak near zero, corresponding to the LiO + TMO phase‑separated region in Figure \ref{fig:GS}b, with the right tail of the $-J_{2,2}/J_{2,1}$ distribution ($> 1$) falls back into the Layered and Spinel‑like region. 
%Although LiO + TMO may raise concerns regarding their suitability as stable single‑phase cathode materials, it does not preclude a theoretical investigation of SRO in their disordered state. 
In the following MC mappings, we still scan through all possible $J_{2,2}/J_{2,1}$ ratios, from Figure \ref{fig:p2p1_PSRO} through Figure \ref{fig:theta}, while only restricting $J_{4,1}$ to near-zero region in the ($J_{2,1},J_{2,2},J_{4,1}$) maps to focus on the space most relevant to LiTMO$_2$ and to reduce the computational cost of this higher‐dimensional parameter space.

\clearpage
\section{ \label{sec: maps} Pair-wise and Four-body Probability Maps: $\alpha_{2,1}$, $\alpha_{2,2}$, and Li${_4}$}
\subsection{\label{sec: p2p1} ($J_{2,1}$, $J_{2,2}$) Maps ($J_{4,1} = 0$)}

\subsubsection{$J_{2,1}$ $>$ 0 and $J_{2,2}/J_{2,1}\in[-2,2]$}
We demonstrate the framework of pair-wise SRO and Li${_4}$ Probability analysis by starting with the simplest yet still important case: Ensemble averaged quantities as functions of $J_{2,1}$ and $J_{2,2}$ interactions only, the two dominant cluster interactions in LiTMO$_2$ (Figure \ref{fig:ECIs}), without $J_{4,1}$ term to break the Layered/Spinel-like degeneracy. 
We first focus on the region where $J_{2,1}$ $>$ 0, $J_{2,2}$/$J_{2,1}$ lies between $\pm$2 (scanned in increments of $J_{2,2}$/$J_{2,1}$ = 0.05), which includes most of the compositions presented in Section \ref{sec: CE}.

\begin{figure}
\centering
\includegraphics[width=0.95\columnwidth]{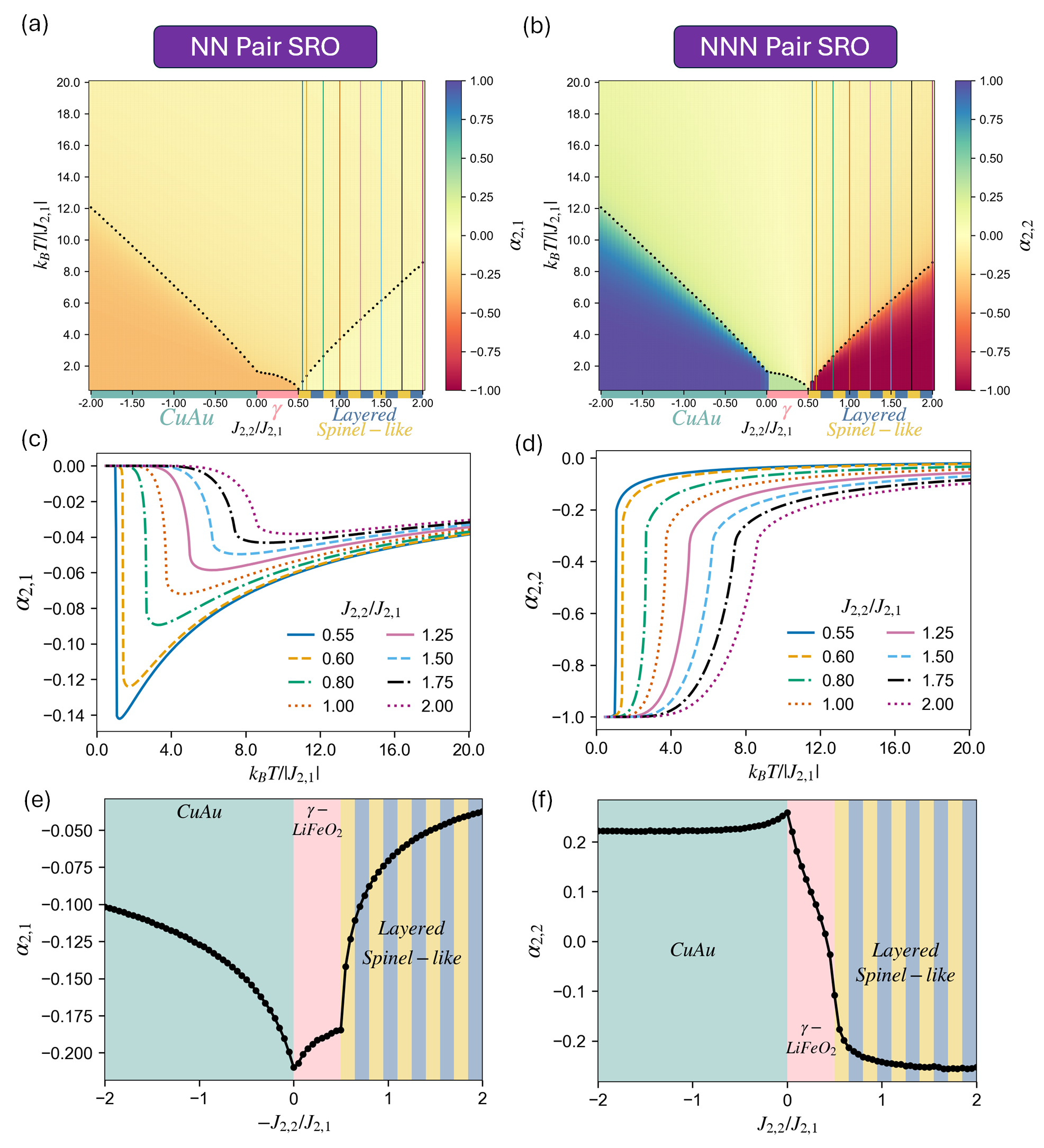}
\caption{\label{fig:p2p1_PSRO} MC simulation results with $J_{2,1} >$ 0, $J_{2,2}$/$J_{2,1}$ between $\pm$2 ($J_{4,1} = 0$) for pair SRO parameters: (a) $\alpha_{2,1}$ and (b) $\alpha_{2,2}$ maps as a function of the $J_{2,2}$/$J_{2,1}$ ratio and temperature. The black dotted curves indicate $T_c$ for each $J_{2,2}$/$J_{2,1}$, defined by the maxima of $\mathrm{Var}(E)/k_{B}T^{2}$. Colored bands beneath the plot mark the stable ground‑state LRO for each $J_{2,2}/J_{2,1}$ interval, with alternating blue and yellow blocks representing Layered/Spinel-like degeneracy. (c) $\alpha_{2,1}$ and (d) $\alpha_{2,2}$ values as a function of temperature at different $J_{2,2}$/$J_{2,1}$ (vertical lines in (a), (b)) demonstrate the SRO trend in systems exhibiting Layered/Spinel-like (degenerate) LRO with $J_{2,1} >$ 0.  (e) $\alpha_{2,1}$ and (f) $\alpha_{2,2}$ quantities at $T/T_c$ = 1.1 are shown across $J_{2,2}$/$J_{2,1}$. 
The background-colored regions and LRO texts classify different $J_{2,2}$/$J_{2,1}$ regions by their corresponding ground state.}
\end{figure}

We begin the analysis by examining $\alpha_{2,1} (= \langle\Gamma_{2,1}\rangle$), the term with the largest coefficient in Equation \ref{eq:Li4} for the Li$_4$ probability, as well as $\alpha_{2,2}$, which does not directly contribute to the Li$_4$ probability but will be shown to affect $\alpha_{2,1}$. 
As shown in Figure \ref{fig:p2p1_PSRO}, $\alpha_{2,1}$ is either zero or negative (mixing), while $\alpha_{2,2}$ transitions from positive (clustering) to negative (mixing) values as $J_{2,2}$ becomes more positive across in both above and below $T_c$. 
This observation is intuitively reasonable since $J_{2,1}$ remains positive (encouraging mixing) for all simulations while we are scanning through $J_{2,2}$ for both positive and negative values. 
As shown in Figure \ref{fig:GS}, each ground-state LRO occupies its own region on the ground-state map corresponding to a specific range of interactions. 
The corresponding correlations and SRO parameters (Figure \ref{fig:p2p1_PSRO}a,b) below $T_c$ remain constant within each LRO region until a discontinuity occurs at $J_{2,2}$/$J_{2,1}$ corresponding to the LRO boundary. 
On the other hand, the SRO above $T_c$ changes continuously across $J_{2,2}$/$J_{2,1}$, although the variation is still more pronounced near the $J_{2,2}$/$J_{2,1}$ at LRO boundary (Figure \ref{fig:p2p1_PSRO}e,f). 
Furthermore, we observe an uncommon trend in $\alpha_{2,1}$ above $T_c$ for Layered/Spinel-like (degenerate) region ($J_{2,2}$/$J_{2,1}$ $>$ 0.5) that breaks the hypothesis of intermediate values and monotonic changes at $\infty>T>T_c$ we previously mentioned.
Starting from $\alpha_{2,1}=0$ at $T=0$ K, the $\alpha_{2,1}$ becomes increasingly negative, away from the random limit, as the temperature rises above $T_c$, but eventually begins trending back toward the $T=\infty$ random limit at higher temperature (Figure \ref{fig:p2p1_PSRO}c). 
In contrast, $\alpha_{2,2}$ follows the expected monotonic trend (Figure \ref{fig:p2p1_PSRO}d), weakening in its magnitude and approaching the random limit with increasing temperature as entropic contributions become more significant.
A similar anomaly is also seen for $J_{2,1} <$ 0 maps and will be further discussed. 
We will explain this abnormal outward curve of $\alpha_{2,1}$ by considering the FCC frustration effects between NN and NNN sites. 
Despite the complex features of the SRO parameters analyzed, the simple fact that $\alpha_{2,1}$ is consistently negative will be shown to be the leading factor of the Li$_4$ deficiency in the following discussion.

\begin{figure}
\centering
\includegraphics[width=0.9\columnwidth]{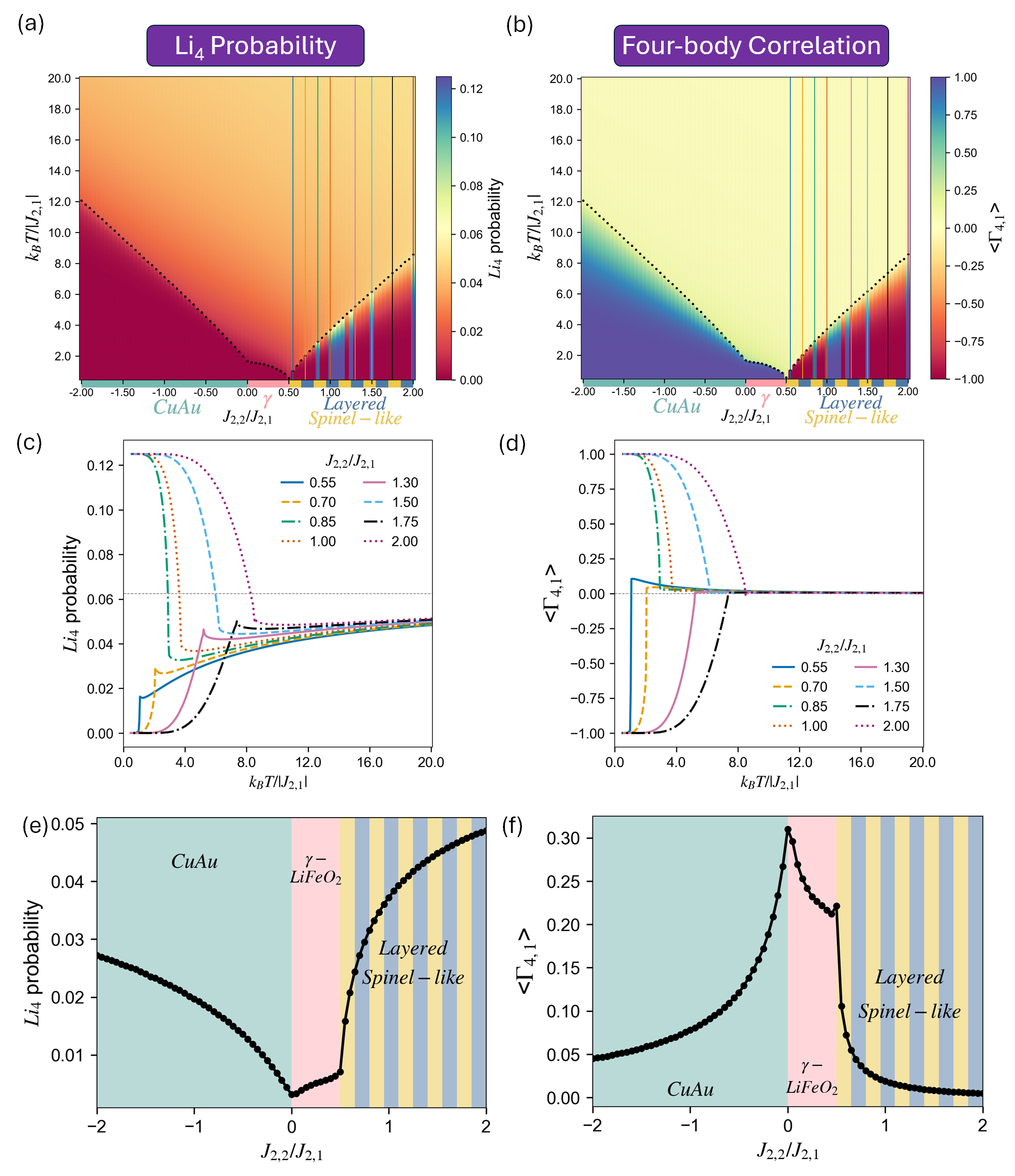}
\caption{\label{fig:p2p1_Li4} MC simulation results with $J_{2,1} >$ 0, $J_{2,2}$/$J_{2,1}$ between $\pm$2 ($J_{4,1} = 0$) for four-body quantities: (a) Li${_4}$ probability and (b) $\langle\Gamma_{4,1}\rangle$ maps as a function of the $J_{2,2}$/$J_{2,1}$ ratio and temperature. $T_c$ are indicated by the black dotted curve. 
%Colored bands beneath the plot mark the stable ground‑state LRO for each $J_{2,2}/J_{2,1}$ interval, with alternating blue and yellow blocks representing Layered/Spinel-like degeneracy. 
(c) Li${_4}$ probability and 
(d) $\langle\Gamma_{4,1}\rangle$ values as a function of temperature at different $J_{2,2}$/$J_{2,1}$ (vertical lines in (a), (b)) in Layered/Spinel-like (degenerate) region with $J_{2,1} >$ 0. 
Please note that all $J_{2,2}$/$J_{2,1}$ in (c) and (d) should have a 50\% probability of equilibrating into either Layered or Spinel-like LRO (as shown in Figure \ref{fig:SI_LvsSP}), while the lines in the plots represent the single result from the run used to build the map.
The corresponding Layered/Spinel-like LRO values for both quantities can be found in Table~\ref{tab:Corr_table}.
(e) Li${_4}$ probability and (f) $\langle\Gamma_{4,1}\rangle$ quantities at $T/T_c$ = 1.1 are shown across $J_{2,2}$/$J_{2,1}$. 
%The background-colored regions and LRO texts classify different $J_{2,2}$/$J_{2,1}$ regions by their corresponding ground state in Figure \ref{fig:GS}a.
}
\end{figure}

Next, we investigate the four-body quantities: the targeted Li${_4}$ probability and the remaining $\langle\Gamma_{4,1}\rangle$ term (four-body correlation) in Equation \ref{eq:Li4}. 
Due to the degeneracy between Layered and Spinel-like orderings before introducing the first four-body interactions ($J_{4,1}$), simulations with $J_{2,2}$/$J_{2,1}$ $>$ 0.5 below $T_c$ have a 50\% probability of equilibrating into either Layered or Spinel-like LRO randomly, as shown in Figure \ref{fig:p2p1_Li4}a,b,c,d and further discussed in the Supporting Information. 
Nevertheless, regardless of the low-temperature LRO, the Li${_4}$ probability in disordered states are consistently below the random limit (0.0625, from $x^4 = (0.5)^4$ at Li:TM = 1:1) across the wide range of $J_{2,2}$/$J_{2,1}$ we scanned. 
This observation illustrates a fundamental point for understanding short-range ordering in the disordered state: the Li$_4$ probability at $\infty>T>T_c$ cannot be consistently inferred from the low-temperature LRO alone, and the hypothesis of intermediate values and monotonic change breaks again; the same holds even after the degeneracy between Layered and Spinel-like structures is lifted by including $J_{4,1}$, as shown later in the $(J_{2,1}, J_{2,2}, J_{4,1})$ maps.

By comparing Figure \ref{fig:p2p1_PSRO}e and Figure \ref{fig:p2p1_Li4}e, we can observe that the Li${_4}$ probability in the disordered state is highly correlated with the ``mixing strength'' from $\alpha_{2,1}$. 
Only below $T_c$, $\langle\Gamma_{4,1}\rangle = \pm 1$ is the dominant term in the equation for the Li$_4$ probability.
At $T>T_c$, $\langle\Gamma_{4,1}\rangle$ has a much smaller magnitude (closer to zero), leaving the $6\,\langle\Gamma_{2,1}\rangle$ = $6\,\alpha_{2,1}$ term in Equation \ref{eq:Li4} to dominate the Li$_4$ probability. 
Consequently, as shown in Figure \ref{fig:p2p1_Li4}c, the curve exhibits a small dip above $T_c$ deviating from the random limit, further reducing the Li$_4$ probability, directly corresponding to the similar dip observed in $\alpha_{2,1}$ at $J_{2,2}$/$J_{2,1}$ $>$ 0.5 above $T_c$. 
The only case where $\langle\Gamma_{4,1}\rangle$ exhibits higher values is in the $\gamma$-LiFeO$_2$ region and CuAu region near $J_{2,2}$/$J_{2,1} \approx$ 0, as the Li$_2$TM$_2$ tetrahedron cluster also contributes positively to $\langle\Gamma_{4,1}\rangle$ value. 
Nevertheless, their most negative $\alpha_{2,1}$, combined with its coefficient factor of 6, corresponds to the lowest Li$_4$ probability in disordered states. 
Results from the simplified $J_{2,2}$/$J_{2,1}$ model support the general qualitative intuition in rocksalt-type LiTMO$_2$: systems with $\gamma$-LiFeO$_2$ groud state have lower Li$_4$ probabilities in the disordered states compared to those with Layered and Spinel-like LRO. 
However, we find that systems exhibiting positive $J_{2,1}$, even those equilibrating into Layered and Spinel-like LRO below $T_c$, consistently present Li$_4$ probabilities at $T=$ 1.1$T_c$ considerably below the random limit.

%%%%%%%%%%%%%%%%%%%%%%%%%%%%%%%%%%%%%%%%%%%%%%%%%%%%%%%%%%%%%%%%%%%%%%%%%%%%%%%%%%%%%%%%%%%%%%%%%%%%%%%%%%%%%%%%%%%%%%%%%%%%%%%%%%%%%%%%%%%%%%%%%%%%%%%%%%%%%%%%%%%%%%%%%%%%%%%%%%%%%%%%%%%%%%%%%%%%%%%%%%%%%%%%%%%%%%%%%%

\subsubsection{$J_{2,1} <$ 0 and $J_{2,2}/J_{2,1}\in[-2,2]$}
\begin{figure}
\centering
\includegraphics[width=0.9\columnwidth]{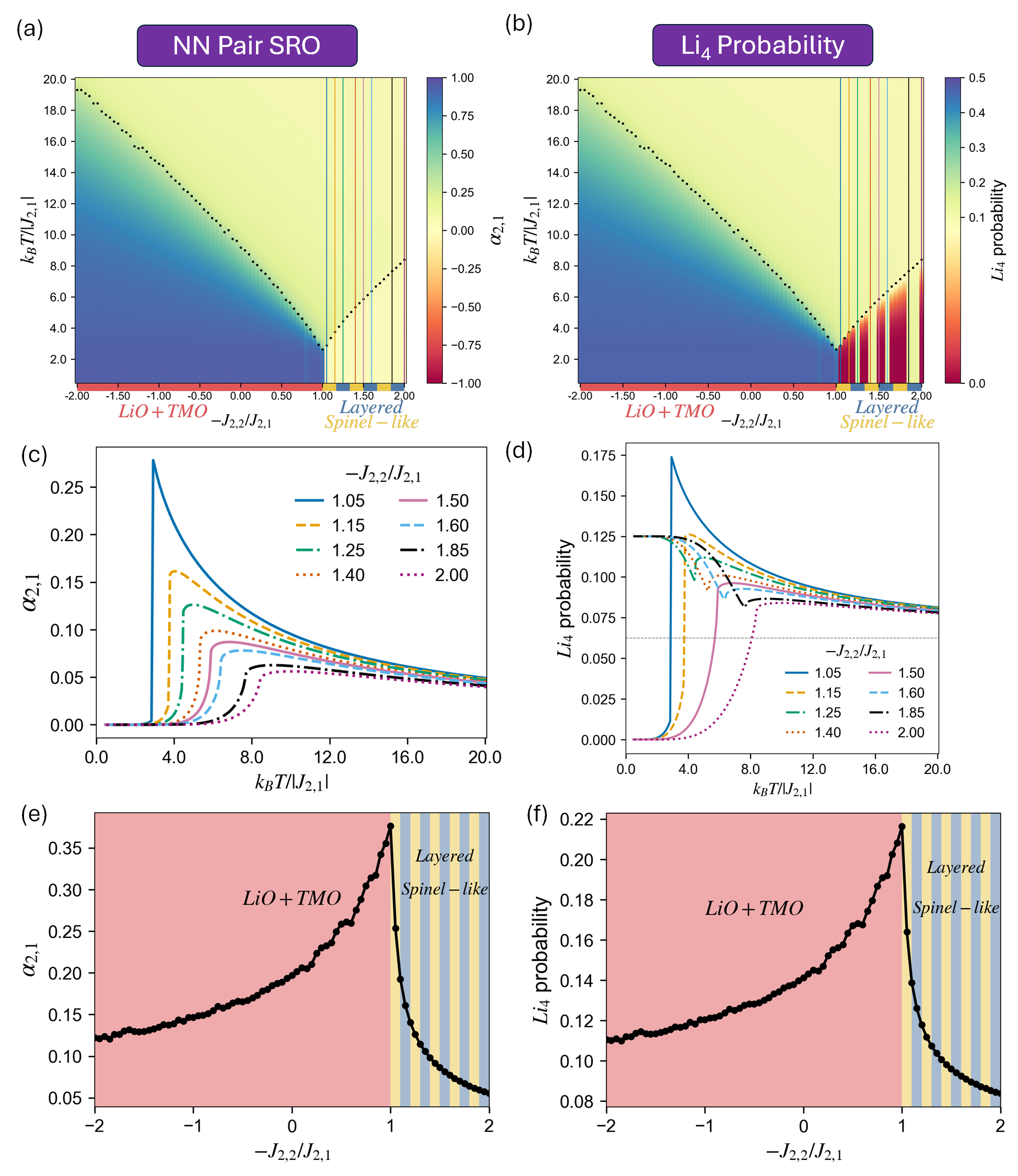}
\caption{\label{fig:Np1} MC simulation results with $J_{2,1} <$ 0, $-J_{2,2}$/$J_{2,1}$ between $\pm$2 ($J_{4,1} = 0$) for (a) $\alpha_{2,1}$ and (b) Li${_4}$ probability maps as a function of the $-J_{2,2}$/$J_{2,1}$ ratio and temperature. $T_c$ are indicated by the black dotted curve. 
%Colored bands beneath the plot mark the stable ground‑state LRO for each $J_{2,2}/J_{2,1}$ interval, with alternating blue and yellow blocks representing Layered/Spinel-like degeneracy. 
Note that, in panel (b), the color bar is centered on the random limit (0.0625) to facilitate comparison with Figure \ref{fig:p2p1_Li4}a. (c) $\alpha_{2,1}$ and (d) Li${_4}$ probability values as a function of temperature at different $-J_{2,2}$/$J_{2,1}$ (vertical lines in (a), (b)) in Layered/Spinel-like (degenerate) region with $J_{2,1} <$ 0. 
Please also note that all $-J_{2,2}$/$J_{2,1}$ in (d) should have a 50\% probability of equilibrating into either Layered or Spinel-like LRO (as shown in Figure \ref{fig:SI_LvsSP}), while the lines in the plots represent the single result from the run used to build the map.
(e) $\alpha_{2,1}$ and (f) Li${_4}$ probability quantities at $T/T_c$ = 1.1 are shown across $-J_{2,2}$/$J_{2,1}$. 
%The background-colored regions and LRO texts classify different $-J_{2,2}$/$J_{2,1}$ regions by their corresponding ground-state LRO in Figure \ref{fig:GS}b.
}
\end{figure}

We next extend our investigation also to the negative $J_{2,1}$ but keep $J_{2,2}/J_{2,1}\in[-2,2]$, which corresponds to the x‑axis of Figure \ref{fig:GS}b with $J_{4,1}$ = 0. Note that all plots in this section employ $-J_{2,2}/J_{2,1}$ as their x‑axis, so that larger positive values of $J_{2,2}$ remain on the right‑hand side of the figure. 
Here, we focus directly on the Li$_4$ probability and its dominant term, $\alpha_{2,1}$.
As shown in Figure \ref{fig:Np1}, both $\alpha_{2,1}$ and Li${_4}$ probability exceed their random limits with $J_{2,1} < 0$, and both quantities reach their maxima at $-J_{2,2}/J_{2,1} = 1$, corresponding to the parameter boundary exhibiting degeneracy between Layered/Spinel-like and phase‑separated LiO + TMO.
Many features observed in the positive $J_{2,1}$ plots also appear here: the quantities exhibit discontinuities below $T_c$ but vary continuously above $T_c$ across $-J_{2,2}/J_{2,1}$, 50\% probability of equilibrating into either Layered or Spinel-like LRO randomly at $-J_{2,2}/J_{2,1} > 1$, and the Li${_4}$ probability is still predominantly governed by $\alpha_{2,1}$. 
Moreover, the uncommon outward (away from the random limit) curve above $T_c$ (Figure \ref{fig:Np1}c,d) persists in both $\alpha_{2,1}$ and Li$_4$ probability with $J_{2,1} < 0$, albeit now in the favorable direction (above random).

\begin{figure}
\includegraphics[width=\columnwidth]{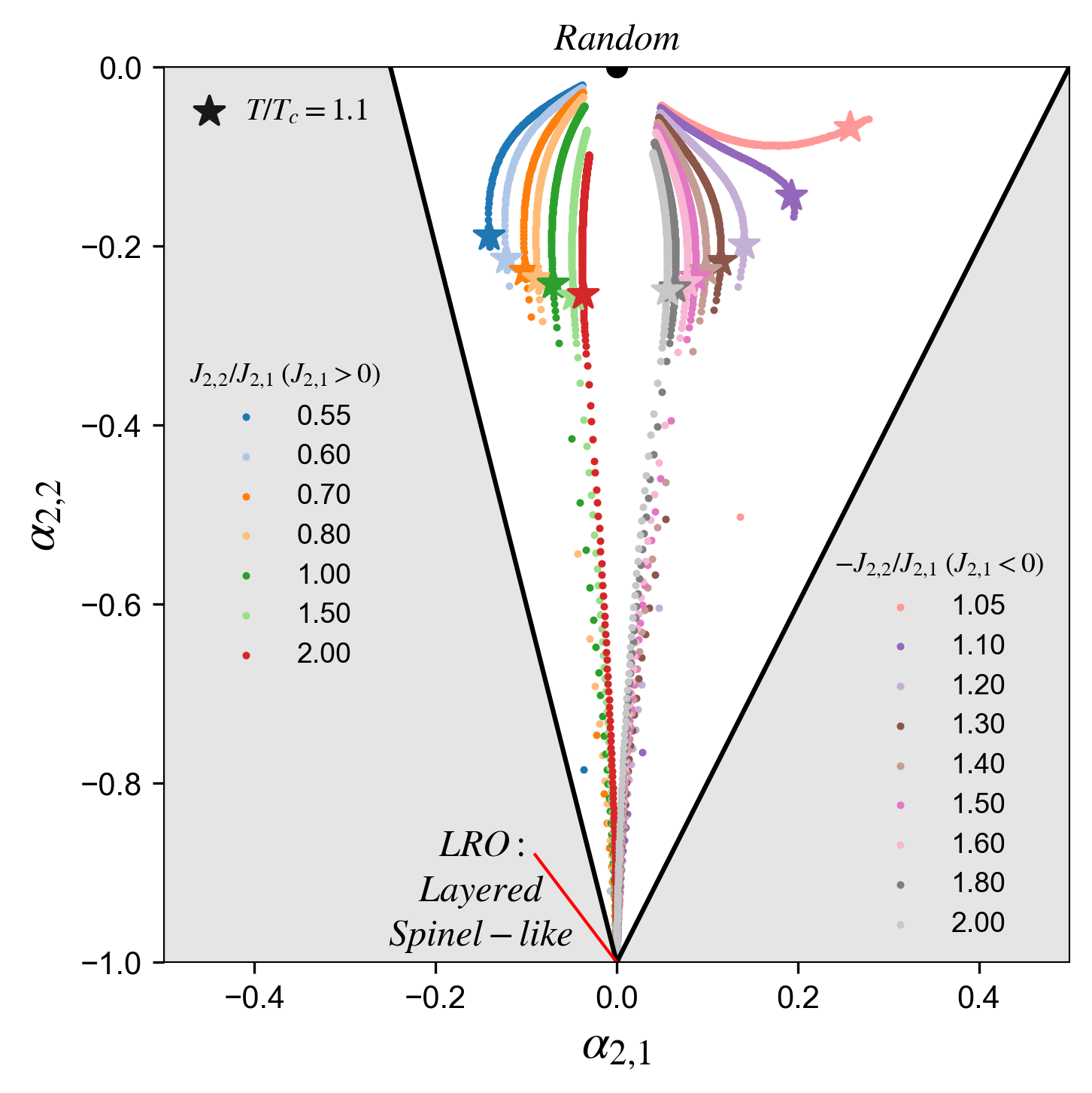}
\caption{\label{fig:alpha} Compilation of $(\alpha_{2,1},\alpha_{2,2})$ pair‑scattering plots across all $k_BT/|J_{2,1}|$ for both $J_{2,1} >$ 0 and $J_{2,1}< 0$ with $(-)J_{2,2}$/$J_{2,1}$ ratio exhibiting Layered and Spinel-like ground-state LRO at vertex (0, -1). The outer thick black lines indicate the region allowed by FCC geometry constraints on NN and NNN sites \cite{rempel1990short, burvsik1999study}.}
\end{figure}

We collect $(\alpha_{2,1},\alpha_{2,2})$ within the Layered/Spinel-like region across all $k_BT/|J_{2,1}|$ from both positive and negative $J_{2,1}$ and plot them in Figure \ref{fig:alpha}. 
Although all scattered curves eventually converge to the vertex (0,–1) as their ground‑state LRO at low temperature, either Layered or Spinel-like, different values of $(-)J_{2,2}$/$J_{2,1}$ generate distinct paths that deviate from the straight line linking (0,–1) to the origin (0,0), which represents the purely random configuration. 
This behavior demonstrates again that, on the FCC lattice, treating pair SRO as merely the residual of its LRO counterpart is not necessarily accurate. 
For both positive and negative $J_{2,1}$, once $J_{2,2}$ exceeds a critical magnitude relative to $J_{2,1}$ (i.e., $J_{2,2} > 0.5J_{2,1}$ for $J_{2,1} > 0$ and $J_{2,2} > -J_{2,1}$ for $J_{2,1} < 0$),  the LRO state prefers keeping all NN pair correlations at their random values 0, sacrificing any energetic benefit from the $J_{2,1}$ term and fully exploiting NNN interactions through perfect NNN sites mixing ($\alpha_{2,2}$ = $\langle\Gamma_{2,2}\rangle=-1$, while NN sites are frustrated in FCC, so $\langle\Gamma_{2,1}\rangle$ has a minimum of $-1/3$ \cite{ducastelle1991order, burvsik1999study, rempel1990short}).
Under perfect NNN mixing, the FCC geometric constraint prevents NN sites from exhibiting any non‑zero correlation (see Figure \ref{fig:str}; opposite occupation on NNN sites forces one same- and one different-type occupation on the two sets of NN sites). 
In disordered states, however, entropy weakens the NNN mixing; with diminished frustration, NN pair correlations can locally fluctuate toward the favorable value encouraged by $J_{2,1}$, thereby modestly lowering the enthalpy, which is the typical SRO phenomenon. 
The uncommon outward curve (away from $\alpha_{2,1} = 0$) can be explained as follows: as temperature rises and the magnitude of $\alpha_{2,2}$ decreases, additional room is created for NN sites to express their preferred local tendency, until sufficiently high temperatures, when entropy fully dominates and both correlations decay toward origin.
We note that visually similar ``evolution paths'' has been reported in past literature during the non-equilibrium evolution over 10,000 MC flips per site at $T/T_c$ = 1/3 for one set of interactions \cite{burvsik1999study}. 
By contrast, we report the ordering paths from equilibrated SRO parameters at 456 temperature points for a single $(-)J_{2,2}$/$J_{2,1}$ ratio using a simulated annealing procedure, with each temperature point produced by 20,000 MC flips per site.

%%%%%%%%%%%%%%%%%%%%%%%%%%%%%%%%%%%%%%%%%%%%%%%%%%%%%%%%%%%%%%%%%%%%%%%%%%%%%%%%%%%%%%%%%%%%%%%%%%%%%%%%%%%%%%%%%%%%%%%%%%%%%%%%%%%%%%%%%%%%%%%%%%%%%%%%%%%%%%%%%%%%%%%%%%%%%%%%%%%%%%%%%%%%%%%%%%%%%%%%%%%%%%%%%%%%%%%%%%

%\subsubsection{\label{sec: theta}%
%  $J_{2,1}=\mathrm{J}\cos\theta,\,J_{2,2}=\mathrm{J}\sin\theta,\,\theta\in[0,2\pi)$
%}

\begin{figure}
\includegraphics[width=\columnwidth]{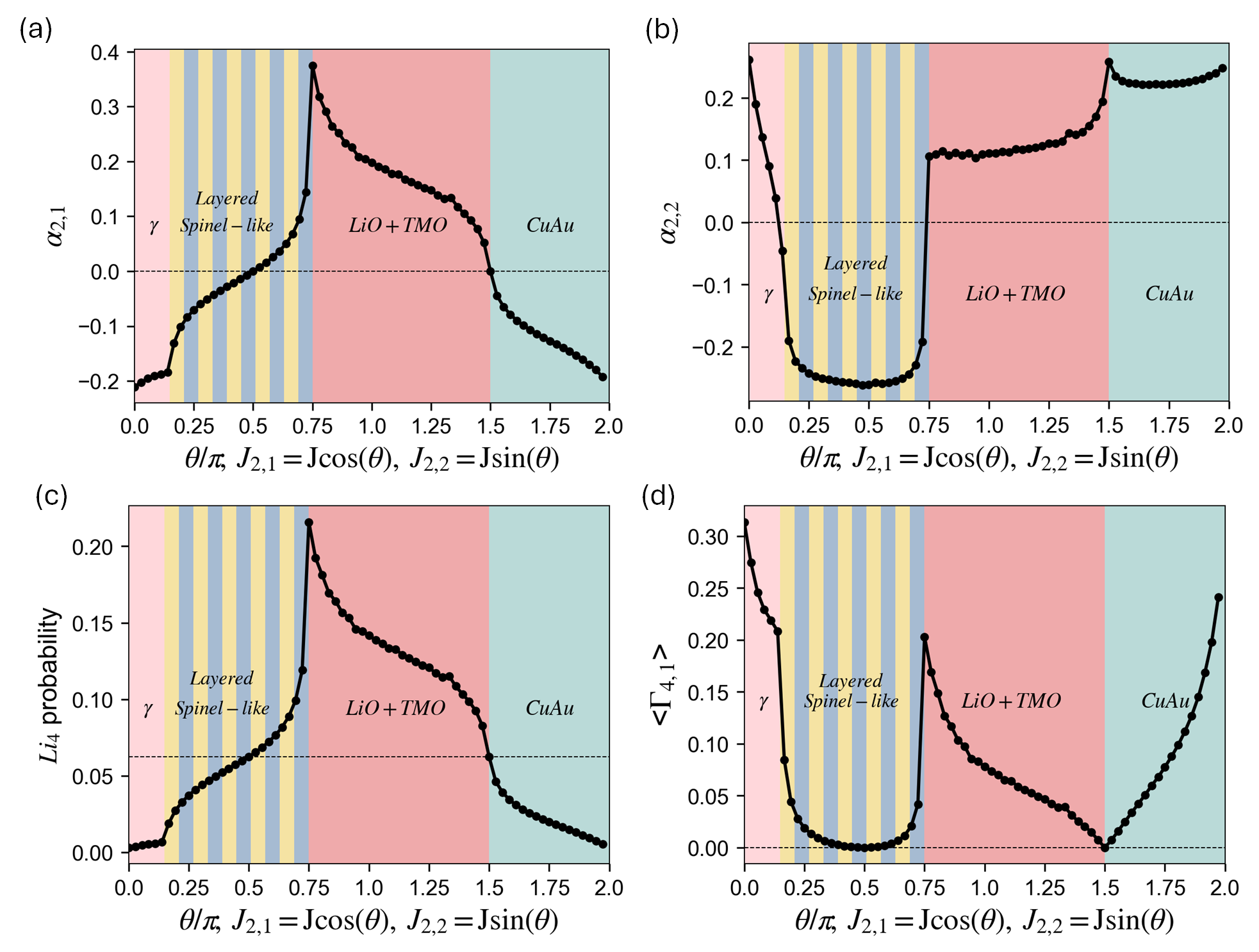}
\caption{\label{fig:theta} MC simulation results at $T/T_c$ = 1.1 with $J_{2,1}=J\cos\theta,\,J_{2,2}=J\sin\theta,\,\theta\in[0,2\pi)$ for (a) $\alpha_{2,1}$, (b) $\alpha_{2,2}$, (c) Li${_4}$ probability, and (d) $\langle\Gamma_{4,1}\rangle$.}
\end{figure}

After discussing the SRO behaviors separately in both positive and negative $J_{2,1}$, we conclude this ($J_{2,1}$, $J_{2,2}$) subsection with a comprehensive scan for all possible $(-)J_{2,2}$/$J_{2,1}$ by parameterizing the interactions as $J_{2,1}=\mathrm{J}\cos\theta,\,J_{2,2}=\mathrm{J}\sin\theta,\,\theta\in[0,2\pi)$, thereby covering both investigated cases and the previously unexamined region (where $J_{2,1}$ close to zero), with all quantities at $T/T_c$ = 1.1 organized in Figure \ref{fig:theta}. 
Within our simplified formalism, whether $\alpha_{2,1}$ and Li${_4}$ probability lie above or below the random limit is entirely determined by the sign of $J_{2,1}$ (i.e., $\theta
\in (\pi/2, 3\pi/2)$ or not). 
As predicted in Figure \ref{fig:OQMD_Hist}, 97\% of our enumerated LiTMO$_2$ (6019 compositions) with positive $J_{2,1}$ fall within the region of $\theta\in (-\pi/2, \pi/2)$, corresponding to the Li${_4}$ probability that is consistently below the random limit.
The shape and the overall trend of the Li${_4}$ probability is predominantly governed by $\alpha_{2,1}$, which is indirectly affected by $\alpha_{2,2}$ via the FCC frustration effects.
When only the first two pair interactions are present, the $\langle\Gamma_{4,1}\rangle$ contribution to the Li${_4}$ probability is negligible compared to $\alpha_{2,1}$, and $\langle\Gamma_{4,1}\rangle$ is always non-negative, dropping to zero when $J_{2,1} = 0$ (at $\theta = \pi/2$ or  $3\pi/2$).
This analysis lays the foundation for our later discussion of strategies to improve the Li${_4}$ probability. 

\clearpage
%%%%%%%%%%%%%%%%%%%%%%%%%%%%%%%%%%%%%%%%%%%%%%%%%%%%%%%%%%%%%%%%%%%%%%%%%%%%%%%%%%%%%%%%%%%%%%%%%%%%%%%%%%%%%%%%%%%%%%%%%%%%%%%%%%%%%%%%%%%%%%%%%%%%%%%%%%%%%%%%%%%%%%%%%%%%%%%%%%%%%%%%%%%%%%

\subsection{($J_{2,1}$, $J_{2,2}$, $J_{4,1}$) Maps ($J_{2,1} > 0$)}

\begin{figure}
\centering
\includegraphics[width=1.0\columnwidth]{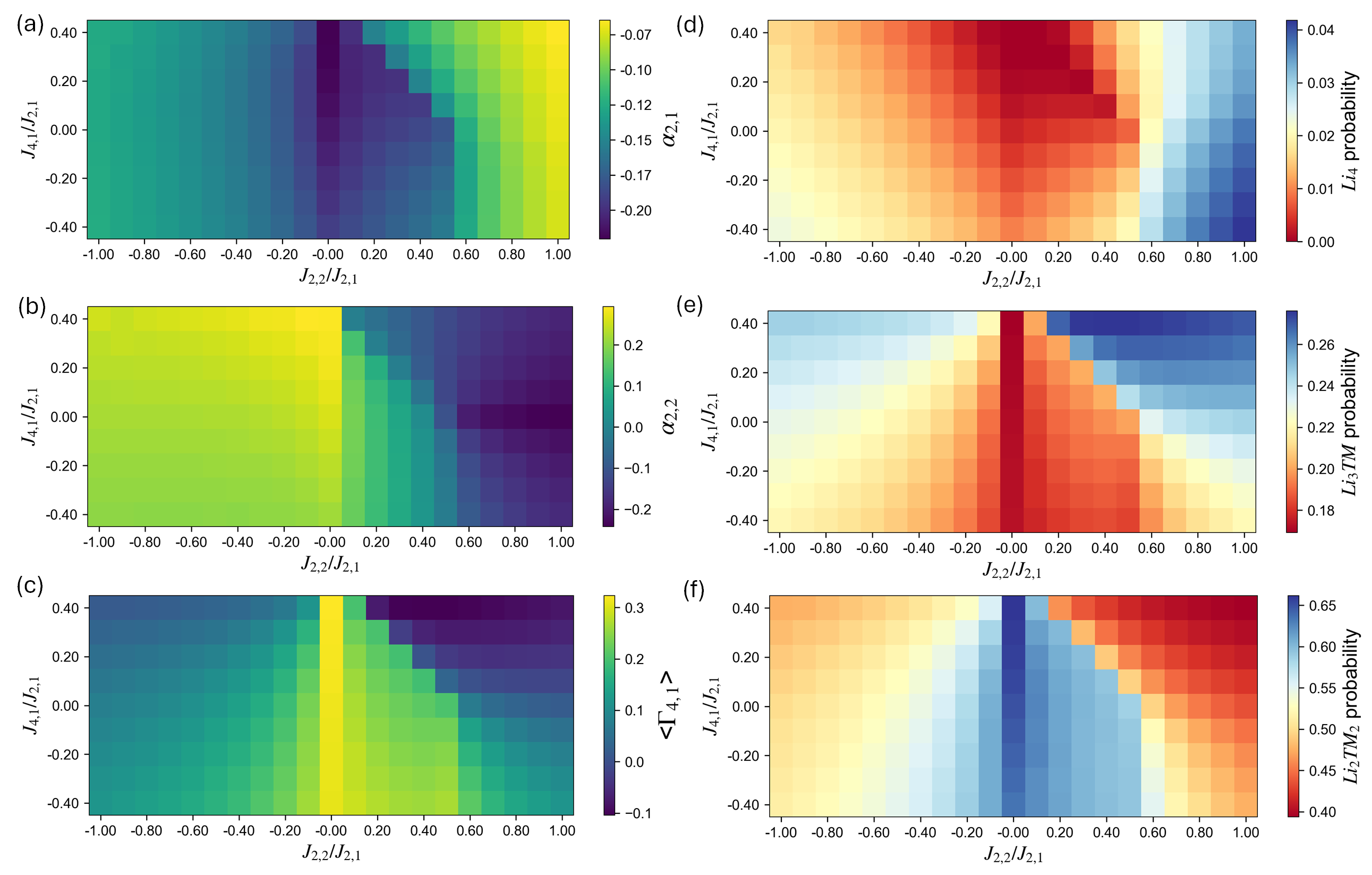}
\caption{\label{fig:Q1P2P1} MC maps for quantities at $T/T_c$ = 1.1 (a) $\alpha_{2,1}$, (b) $\alpha_{2,2}$, (c)  $\langle\Gamma_{4,1}\rangle$, and (d) Li${_4}$, (e) Li${_3}$TM${_1}$, (f) Li${_2}$TM${_2}$ tetrahedron cluster probability as a function of $J_{2,2}$/$J_{2,1}$ and $J_{4,1}$/$J_{2,1}$ ($J_{2,1} > 0$).}
\end{figure}

We next extend our investigation to include $J_{4,1}$. This interaction that assigns opposite energy contributions to tetrahedra with an even number of (Li${_4}$, Li${_2}$TM${_2}$, TM${_4}$; contributing $\langle\Gamma_{4,1}\rangle = 1$) and those with an odd number of Li (Li${_3}$TM, Li${_3}$TM; contributing $\langle\Gamma_{4,1}\rangle = -1$). 
In our approximate projected ECIs formulation (Equation \ref{DRX_inv}), this term directly reflects the energy difference between the Layered and Spinel-like structures.
In this section, we focus on Li$_x$TM$_{4-x}$ probabilities.
We again note the analytical connection between Li$_x$TM$_{4-x}$ probabilities and their subcluster correlations (Equation \ref{C_matrix}), and all of these quantities are shown in Figure \ref{fig:Q1P2P1}.
SRO quantities in the disordered state exhibit continuous variation across most of the interaction parameter space, except for discrete jumps near the $\gamma$-LiFeO$_2$ and Layered between ($J_{2,2}$/$J_{2,1}$, $J_{4,1}$/$J_{2,1}$) = (0.5, 0) and (0, 0.5); the sharpest discontinuity occurs in $\langle\Gamma_{4,1}\rangle$ near (0, 0.5).
The pronounced transitions on the maps coincide with the same boundary positions as in the ground-state LRO diagram (Figure \ref{fig:GS}a), similar to the observations made in ($J_{2,1}$, $J_{2,2}$) maps. 
However, in Layered and Spinel-like regions, absolute values of correlations and tetrahedron cluster probabilities are again not simply attenuated versions of the corresponding LRO values (Table \ref{tab:Corr_table}).
Across the window of interactions we scanned, $\alpha_{2,1}$ is consistently negative, the Li$_4$ probability remains below two-thirds of the random limit, and the Li$_2$TM$_2$ probability remains above the random limit. 
From Equation \ref{C_matrix}, the high Li$_2$TM$_2$ probability is a natural result of the negative $\alpha_{2,1}$ at $\infty>T>T_c$, even for systems exhibiting Layered and Spinel-like ground states that are constrained to $\alpha_{2,1} =0$ at $T=0$ K.
Nevertheless, qualitative links can be drawn between Li$_x$TM$_{4-x}$ trends in the ($J_{2,1}$, $J_{2,2}$, $J_{4,1}$) SRO maps and those in the ground-state LRO diagram.

By introducing $J_{4,1}$ to break the degeneracy between Layered and Spinel-like LRO, we find that increasingly negative $J_{4,1}$, which favors the Spinel-like LRO (featuring Li$_4$ probability two times higher than the random limit) in the ground-state diagram, raises the Li$_4$ probability in the disordered state. 
On the other hand, increasingly positive $J_{4,1}$, which favors the Layered (featuring P(Li$_3$TM) = P(LiTM$_3$) = 0.5) LRO, correspondingly increases the Li$_3$TM probability.
Such changes can be considered in terms of the variations in $\langle\Gamma_{4,1}\rangle$. For instance, from ($J_{2,2}$/$J_{2,1}$, $J_{4,1}$/$J_{2,1}$) = (1, 0.4) to (1, -0.4), $\langle\Gamma_{4,1}\rangle$ increases from -0.08 to 0.12. 
However, the Li$_4$ probability remains largely affected by the $\alpha_{2,1}$ contributions (Equation \ref{eq:Li4}), which even exhibit weakly opposite dependence on $J_{4,1}$ relative to $\langle\Gamma_{4,1}\rangle$ at $J_{2,2}$/$J_{2,1} > 0.5$ region, decreasing the efficiency of improving Li$_4$ probability via $J_{4,1}$. 
From ($J_{2,2}$/$J_{2,1}$, $J_{4,1}$/$J_{2,1}$) = (1, 0.4) to (1,-0.4), Li$_4$ probability only increases from 0.033 to 0.042. For comparison, from ($J_{2,2}$/$J_{2,1}$, $J_{4,1}$/$J_{2,1}$) = (0.6, 0) to (1, 0), Li$_4$ probability increases from 0.021 to 0.037.
When combined with the narrow $J_{4,1}/J_{2,1}$ distribution characteristic of most LiTMO$_2$ compositions, pursuing chemistry that strongly favor a spinel-like LRO, though still helpful, does not necessarily yield substantial improvements in Li$_4$ probability of the fully disordered phase. 
For chemistry favors Spinel-like ordering, it is better to target partially or fully LRO states below $T_c$ \cite{cai2024situ}. 
If a fully disordered state is the goal, then for most LiTMO$_2$ compositions with positive $J_{2,1}$, maintaining a sufficiently large $J_{2,2}/J_{2,1}$ ratio appears to be an important prerequisite for achieving a high Li$_4$ probability.

%%%%%%%%%%%%%%%%%%%%%%%%%%%%%%%%%%%%%%%%%%%%%%%%%%%%%%%%%%%%%%%%%%%%%%%%%%%%%%%%%%%%%%%%%%%%%%%%%%%%%%%%%%%%%%%%%%%%%%%%%%%%%%%%%%%%%%%%%%%%%%%%%%%%%%%%%%%%%%%%%%%%%%%%%%%%%%%%%%%%%%%%%%%%%%

\section{\label{sec: discussion} Discussion: Improving Li$_4$ Probability}

In this section, we provide extra insights into achieving a high Li$_4$ probability in the disordered state based on our MC maps.
First, for most compositions exhibiting positive $J_{2,1}$, one should expect the Li$_4$ probability to be lower than random limit regardless of ground-state LRO. 
To increase the Li$_4$ probability in this category, the most straightforward approach is to maximize $J_{2,2}$/$J_{2,1}$ by selecting chemistry that strongly favor Layered or Spinel-like LRO. 
Moving further toward the Spinel-like side, which lowers $J_{4,1}$/$J_{2,1}$, offers an additional but small benefit in disordered states.

Next, Li$_4$ probability can approach or exceed the random limit by tuning pair interactions (to approach or reach the negative $J_{2,1}$ region in Figure \ref{fig:theta}) and the corresponding SRO with selected compositions.
For reasons of synthesizability and single-phase stability, compositions whose ground state is LiO + TMO phase separation might be avoided. 
In this case, favorable SRO can still be achieved by selecting chemistry that fall deep within the Layered or Spinel-like ground-state region (in the $\theta$ scheme of Figure \ref{fig:theta}, $ 0.5 \leq \theta \leq$ 0.75 ), far away from its boundary with $\gamma$-LiFeO$_2$ LRO, to the extent of discouraging mixing at NN sites and promoting mixing at NNN sites, regardless of the specific physical mechanism involved.
Specifically, we can again leverage the approximate projected-ECI formulation of Equation \ref{DRX_inv} to identify the characteristics that compositions in this category might exhibit.
Based on observations, for $J_{2,2}$ to keep positive while holding $J_{2,1}$ negative or near zero, and if $E_{\mathrm{Layered}}$ and $E_{\mathrm{Spinel\text{-}like}}$ are nearly degenerate and both lower than $E_{\mathrm{Random}}$, then $E_{\gamma\text{-LiFeO}_2} > E_{\mathrm{Random}}$ should be satisfied. 
This result can be intuitively understood: for compositions exhibiting Spinel-like or Layered ordering tendencies, a necessary requirement for Li$_4$ probability to surpass the random limit in their high-temperature disordered states is that their $\gamma$-LiFeO$_2$ ordering energetics be less favorable than the random limit. 
Further quantitative relationships among the ordering energetics may be obtained by solving the system of inequalities derived from Equation \ref{DRX_inv}, while keeping in mind the approximate nature of the projected-ECI formulation.
To demonstrate such a composition, we present two examples from the OQMD: Li$_4$TiVCo$_2$O$_8$ with $E_{\mathrm{Random}}$ = 78 meV/atom, $E_{\mathrm{Layered}}$ = 34 meV/atom $E_{\mathrm{Spinel\text{-}like}}$ = 41 meV/atom, $E_{\gamma\text{-LiFeO}_2}$ = 93 meV/atom, and Li$_4$Co$_2$RuRhO$_8$ with $E_{\mathrm{Random}}$ = 105 meV/atom, $E_{\mathrm{Layered}}$ = $E_{\mathrm{Spinel\text{-}like}}$ = 42 meV/atom, $E_{\gamma\text{-LiFeO}_2}$ = 149 meV/atom. Their projected $J_{2,2}/J_{2,1}$ ratios are 16.9 ($J_{2,1}$ near zero) and  -1.8 (negative $J_{2,1}$), respectively. If these compositions are well described by our pseudobinary, short-range approximations and projected-ECI estimations, they might exhibit Li$_4$ probabilities that approach or surpass the random limit while maintaining single-phase stability at a reasonable synthesis temperature.
Finally, the Li$_4$ probability depends on their equilibrated temperature. 
According to the curves in Figure \ref{fig:p2p1_Li4} and \ref{fig:Np1} corresponding to each system’s interactions, the non-monotonic behavior of Li$_4$ probability can be exploited for fine-tuning.
For most compositions with positive $J_{2,1}$, this recommendation implies staying just above $T_c$, or at a much higher temperature that passes the dip in the curves. 

Beyond the above strategies, additional complexities ignored in our simplified framework, not necessarily enhancing Li$_4$ probability, should be considered.
Firstly, if systems exhibit long-range contributions to the ordering energetics that outweigh those captured by the short-range clusters, additional interaction terms must be incorporated into the formalism for SRO determinations.
Secondly, when multiple TM cations are present, strong mixing or repulsive interactions among TM species can break Li–TM pseudobinary approximations and require extension of the current binary framework to capture the higher-dimensional nature of multi-component systems.
Lastly, the preceding conclusions completely assume thermodynamic equilibrium. 
Kinetic factors must therefore be considered separately to interpret experimental results. 
Nevertheless, the ordering phenomena predicted here by basic statistical mechanics provide a baseline for the fundamental thermodynamic understanding of SRO behaviors arising from the commonly dominant cluster interactions. 

%%%%%%%%%%%%%%%%%%%%%%%%%%%%%%%%%%%%%%%%%%%%%%%%%%%%%%%%%%%%%%%%%%%%%%%%%%%%%%%%%%%%%%%%%%%%%%%%%%%%%%%%%%%%%%%%%%%%%%%%%%%%%%%%%%%%%%%%%%%%%%%%%%%%%%%%%%%%%%%%%%%%%%%%%%%%%%%%%%%%%%%%%%%%%%
\section{\label{sec: conclusion}  Conclusion}

In this study, we conduct comprehensive MC mapping to elucidate the SRO behavior underlying the observed Li$_4$ deficiency in DRX and to formulate strategies for improving the Li$_4$ probability. 
Building upon the significant energetic contributions of short-range clusters in LiTMO$_2$, we employ simplified CE formulations that enable an exhaustive scan of the cluster-interaction parameter space. 
Our results demonstrate 
(1) The major contributions of NN pair SRO on the Li$_4$ probability in the disordered states. 
(2) Li$_4$ probability and NN pair SRO above or below the random limit are primarily determined by the sign of the NN pair interaction.
(3) Ordering quantities above $T_c$ cannot always be safely regarded as ``remnant/precursors'' of the LRO below $T_c$ in FCC geometry.
(4) Tuning $J_{2,2}/J_{2,1}$ and $J_{4,1}/J_{2,1}$ through compositional design can be exploited for improving Li$_4$ probability. 
This research advances the basic knowledge of FCC ordering phenomena while demonstrating how targeted control of SRO may mitigate or even reverse its commonly observed adverse effects in DRX.

% Acknowledgements
\medskip
\textbf{Acknowledgements} \par 
T.-c.L. and C.W. acknowledges funding from Toyota Research Institute. T.-c.L. also acknowledges the support from Taiwanese Government Fellowship for Overseas Study. The authors acknowledge computational resources from the Quest high-performance computing facility at Northwestern University, which is jointly supported by the Office of the Provost, the Office for Research, and Northwestern University Information Technology.

\bibliographystyle{MSP}
\bibliography{ref}

\clearpage
%--------------------------------------------------------------------%
%--------------------------------------------------------------------%
\section*{Supporting Information}
%--------------------------------------------------------------------%
%--------------------------------------------------------------------%

\subsection*{Derivation of Configuration Matrix}

The configuration matrix (C-matrix) shown in Equation~\ref{C_matrix} can be derived by first writing out the contribution of each tetrahedron configuration to each subcluster correlation:

{\footnotesize
\begin{equation}\label{eq:FullMatrix}
\begin{pmatrix}
\langle\Gamma_{0000}\rangle \\[3pt]\hline
\langle\Gamma_{1000}\rangle \\
\langle\Gamma_{0100}\rangle \\
\langle\Gamma_{0010}\rangle \\
\langle\Gamma_{0001}\rangle \\[3pt]\hline
\langle\Gamma_{1100}\rangle \\
\langle\Gamma_{1010}\rangle \\
\langle\Gamma_{1001}\rangle \\
\langle\Gamma_{0110}\rangle \\
\langle\Gamma_{0101}\rangle \\
\langle\Gamma_{0011}\rangle \\[3pt]\hline
\langle\Gamma_{1110}\rangle \\
\langle\Gamma_{1101}\rangle \\
\langle\Gamma_{1011}\rangle \\
\langle\Gamma_{0111}\rangle \\[3pt]\hline
\langle\Gamma_{1111}\rangle \\
\end{pmatrix}
=
\left(
\begin{array}{c|cccc|cccccc|cccc|c}

+ & + & + & + & + & + & + & + & + & + & + & + & + & + & + & + \\[3pt]\hline
+ & - & + & + & + & - & - & - & + & + & + & - & - & - & + & - \\
+ & + & - & + & + & - & + & + & - & - & + & - & - & + & - & - \\
+ & + & + & - & + & + & - & + & - & + & - & - & + & - & - & - \\
+ & + & + & + & - & + & + & - & + & - & - & + & - & - & - & - \\[3pt]\hline
+ & - & - & + & + & + & - & - & - & - & + & + & + & - & - & + \\
+ & - & + & - & + & - & + & - & - & + & - & + & - & + & - & + \\
+ & - & + & + & - & - & - & + & + & - & - & - & + & + & - & + \\
+ & + & - & - & + & - & - & + & + & - & - & + & - & - & + & + \\
+ & + & - & + & - & - & + & - & - & + & - & - & + & - & + & + \\
+ & + & + & - & - & + & - & - & - & - & + & - & - & + & + & + \\[3pt]\hline
+ & - & - & - & + & + & + & - & + & - & - & - & + & + & + & - \\
+ & - & - & + & - & + & - & + & - & + & - & + & - & + & + & - \\
+ & - & + & - & - & - & + & + & - & - & + & + & + & - & + & - \\
+ & + & - & - & - & - & - & - & + & + & + & + & + & + & - & - \\[3pt]\hline
+ & - & - & - & - & + & + & + & + & + & + & - & - & - & - & +
\end{array}
\right)
\begin{pmatrix}
P(++++) \\[3pt]\hline
P(-+++) \\
P(+-++) \\
P(++-+) \\
P(+++-) \\[3pt]\hline
P(--++) \\
P(-+-+) \\
P(-++-) \\
P(+--+) \\
P(+-+-) \\
P(++--) \\[3pt]\hline
P(---+) \\
P(--+-) \\
P(-+--) \\
P(+---) \\[3pt]\hline
P(----) \\
\end{pmatrix}.
\end{equation}}\\
Here, ``+'' and ``-'' denote +1 and -1, respectively.  
The left-hand side column vector lists each subcluster correlation $\langle\Gamma_{\alpha_1 \alpha_2 \alpha_3 \alpha_4}\rangle$, 
where $\alpha_i$ is 1 if site $i$ ($i=1,\dots,4$) in the tetrahedron belongs to the cluster $\alpha$ and 0 otherwise.
The right-hand side column vector lists probabilities P($\sigma_1\sigma_2\sigma_3\sigma_4$) for all 16 possible occupation combinations on four sites in an tetrahedron, where $\sigma_i$ is the occupation variable introduced in Equation~\ref{CE}.
The matrix in the middle records the cluster correlation values (i.e. the products of occupation variables) for each tetrahedron configuration.
The probability‐weighted sum of these subcluster correlations over all tetrahedral configurations provides an alternative way to compute the overall averaged values for the entire supercell configuration. 
Both the cluster indices $\alpha_i$ and the occupation variables $\sigma_1$ follow the naming convention of ATAT \cite{van2002self, van2009multicomponent}. 

To reach Equation~\ref{C_matrix}, we invert the relationship so that the probability column vector appears on the left-hand side.  
Since the central matrix is orthogonal and symmetric (orthonormal when scaled by $1/4$), its inverse is simply itself multiplied by $1/16$, which is the coefficient appearing in Equation~\ref{C_matrix}.  
The final step is to group all equivalent subcluster correlations and tetrahedron probabilities into the reduced column vectors shown in Equation~\ref{C_matrix}.  
Summing all of the +1 and -1  in each block (separated by the auxiliary lines) yields the coefficients of the reduced coefficient matrix, which is the final C-matrix used in this work.

\subsection*{Validation of the Short-Range Cluster Approximation}

To validate the effectiveness of short-range cluster approximation in capturing cluster energetics and targeted ordering parameters, we compare the standard CE of LiTMO$_2$ (Figure \ref{fig:ECIs}) with the fitting the simplified CE that includes only short-range clusters, namely the first two pair clusters and the first four-body cluster, in Table \ref{tab:CE_parameters}. 
We found that fitted ECIs remained similar between two fittings, suggesting that the approximation retains the correct energetic projections for the energetically dominant short-range clusters.
Excluding long-range clusters as additional fitting parameters worsens the CV score, increasing it slightly in the Cu system and by up to a factor of three in the Cr system; on average, the CV score is approximately twice as high across all systems with the short-range approximation, revealing limitations in accurately fitting many small ordered structures.

We then select TM = Cu and Cr systems for further validation of the key ordering parameters, $\alpha_{2,1}$ and Li$_4$ probability in the disordered states at $T/T_c$ = 1.1, to evaluate how the poorer fit affects ordering parameters.
MC simulations were performed with a simulation cell of 32$^3$ sites, with 10,000 MC flips per site for both equilibration and averaging. 
For Cu, $\alpha_{2,1}$ values are -0.058 and -0.063, and Li$_4$ probabilities are 0.043 and 0.039 for the standard and simplified CE, respectively.
The good performance is not surprising given the similar quality of both fittings.
Nevertheless, for Cr, $\alpha_{2,1}$ values are -0.079 and -0.083, and Li$_4$ probabilities are 0.037 and 0.031 for the standard and simplified CE, respectively.
Predictions of high‑temperature short-range ordering parameters from the simplified CE remain sufficiently precise, particularly for the primary purpose of comparing with the random‑limit Li$_4$ probability (0.0625).
The small discrepancy of approximately 0.005 (less than 10\% of the random limit) in Li$_4$ probabilities can be primarily attributed to the three-body ECI and the correlation $\langle\Gamma_{3,1}\rangle$ (with a coefficient of -4/16 in Equation~\ref{eq:Li4}), which is included in the standard CE but omitted from our simplified CE.
Analyzing general ordering behavior in the disordered state, rather than constructing a precise Hamiltonian for each LiTMO$_2$, is the primary focus of this study, and the results above confirm that our approximations are effective for this purpose.

\begin{table}[ht]
\centering
\caption{Comparison between the standard and simplified CE fittings for LiTMO$_2$ (TM = Cr, Cu, Fe, Mn, Ni, and Ti).}
\begin{tabular}{lcccccc}
\toprule
& \textbf{Cr} & \textbf{Cu} & \textbf{Fe} & \textbf{Mn} & \textbf{Ni} & \textbf{Ti} \\
\midrule
Number of structures& \textbf{115} & \textbf{97} & \textbf{111} & \textbf{113} & \textbf{110} & \textbf{112} \\
\midrule

\multicolumn{7}{l}{\textbf{Standard Cluster Expansion Fitting}} \\
$J_{0}$                  & -0.026 & 0.000 & 0.000 & -0.016 & -0.011 & -0.005 \\
$J_{2,1}$                &  0.136 & 0.048 & 0.085 &  0.092 &  0.037 &  0.098 \\
$J_{2,2}$                &  0.127 & 0.056 & 0.082 &  0.056 &  0.049 &  0.052 \\
$J_{4,1}$                &  0.000 & 0.000 & -0.003 &  0.000 &  0.000 &  0.000 \\
10-fold CV (meV/site)       & 18     & 25    & 15    & 26     & 8      & 21     \\
\midrule
\multicolumn{7}{l}{\textbf{ Fitting Using Only Short-range Clusters}} \\
$J_{0}$                  & -0.012 & -0.001 & 0.007 & -0.003 & -0.002 & 0.005 \\
$J_{2,1}$                &  0.137 & 0.048 & 0.077 &  0.093 &  0.037 & 0.098 \\
$J_{2,2}$                &  0.116 & 0.055 & 0.069 &  0.048 &  0.046 & 0.045 \\
$J_{4,1}$                &  0.009 & 0.002 & -0.002 & 0.003 & 0.002 & -0.003 \\
10-fold CV (meV/site)          & 54     & 29    & 37    & 46     & 19     & 60     \\
\bottomrule
\end{tabular}
\label{tab:CE_parameters}
\end{table}

\subsection*{Assessing Projected ECIs as Sufficient Proxies for the ECI Distribution Estimation of LiTMO$_2$}

\begin{figure}
\centering
\includegraphics[width=0.8\columnwidth]{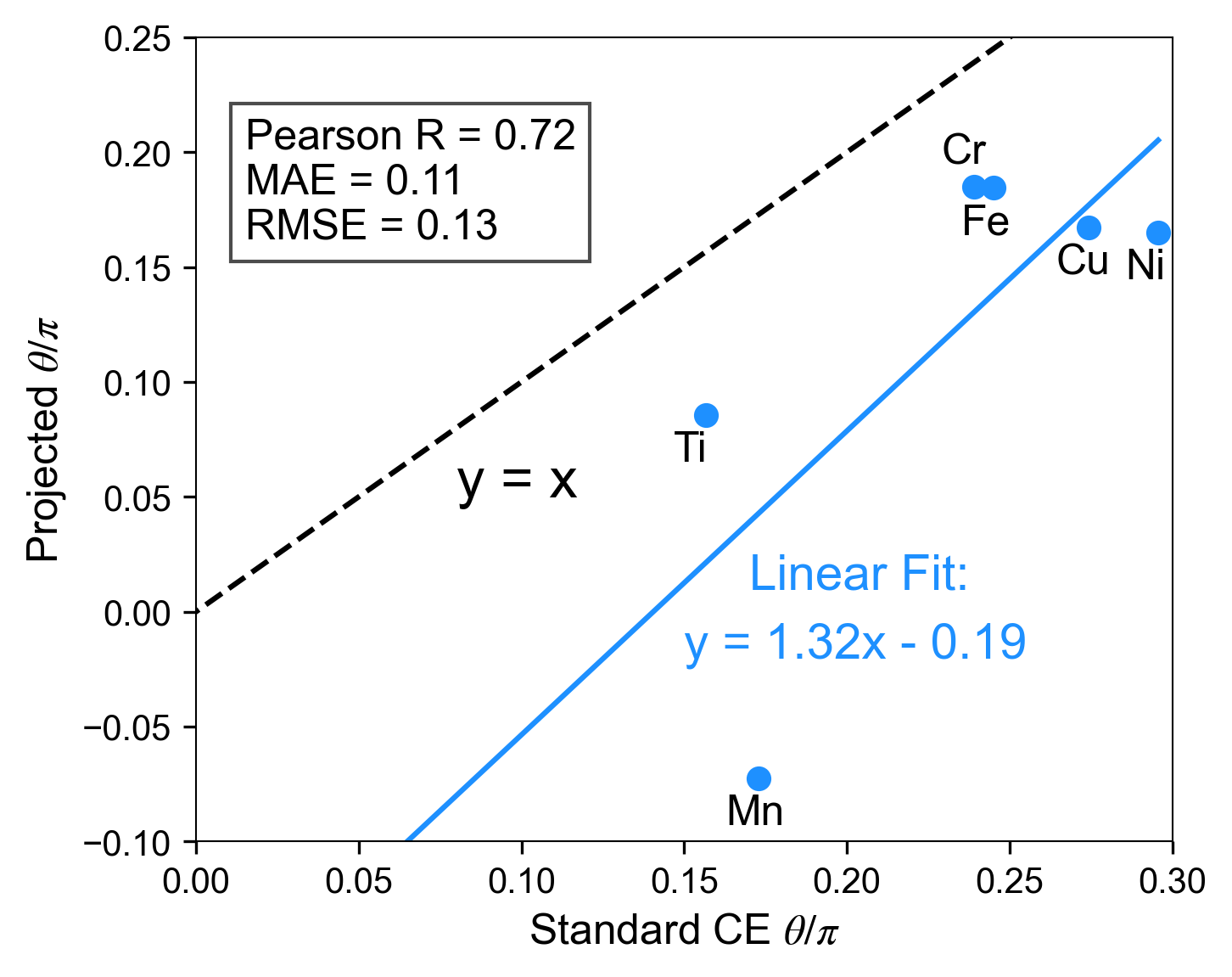}
\caption{\label{fig:SI_pECI} 
Scattering plot of the $\theta /\pi,\ \theta = \tan^{-1}(\frac{J_{2,2}}{J_{2,1}}), J_{2,1}>0$ obtained from the standard CE (Figure \ref{fig:ECIs}) and from the projected scheme (Equation \ref{DRX_inv}). 
Across all chemistries, $J_{2,1}$ is positive in both approaches, which supports one of the central conclusion of this work that positive $J_{2,1}$ ($\theta\in (-\pi/2, \pi/2)$ in Figure \ref{fig:theta}) is the leading cause of the predominantly observed Li$_4$ probabilities lower than the random limit.
With a positive correlation and errors below the level that could move most systems around or below $\theta/\pi = 0.25$ ($J_{2,2}/J_{2,1} = 1$) beyond $\theta/\pi = 0.5$, the relevant ECI parameter space can be safely defined, while a precise one-to-one correspondence is neither expected nor required in this work.
}
\end{figure}

Figure \ref{fig:OQMD_Hist} presents the estimated distribution of key ECIs for LiTMO$_2$, which, when translated into the $\theta$ parameterization in Figure \ref{fig:theta}, explains the predominately observed Li$_4$ probabilities below the random limit for most LiTMO$_2$ compositions.
This section provides evidence that such projection based on only four ordering energies (Equation \ref{DRX_inv}), rather than the hundreds of training-structure energies used in the standard CE, can still yield a reasonable estimate of $\theta$ that governs the ordering parameters in MC maps.
Figure \ref{fig:SI_pECI} reveals the positive correlation of projected ECIs and the true $\theta = \tan^{-1}(\frac{J_{2,2}}{J_{2,1}})$ from the standard CE, and all the positive $J_{2,1}$ ($\theta\in (-\pi/2, \pi/2)$) are correctly captured by the projected ECI scheme.  
Moreover, a deviation of $\theta /\pi$ around 0.1 has negligible impact on understanding the ordering trend across $\theta\in[0,2\pi)$ in Figure \ref{fig:theta}.
This observation supports the effectiveness of the projected ECI in estimating the parameter space of interest for LiTMO$_2$ at a holistic level across $\theta\in[0,2\pi)$. 
The error in the projected ECI scheme serves as a reminder of its approximate nature, particularly its consistent underestimation of the normalized $J_{2,2}$ strength relative to that from the standard CE.
As noted in the main text, Equation \ref{DRX_inv} is neither expected nor intended to replace the standard CE fitting method (otherwise, there would be no point in using the much more expensive procedure).

\clearpage

\begin{figure}
\centering
\includegraphics[width=\columnwidth]{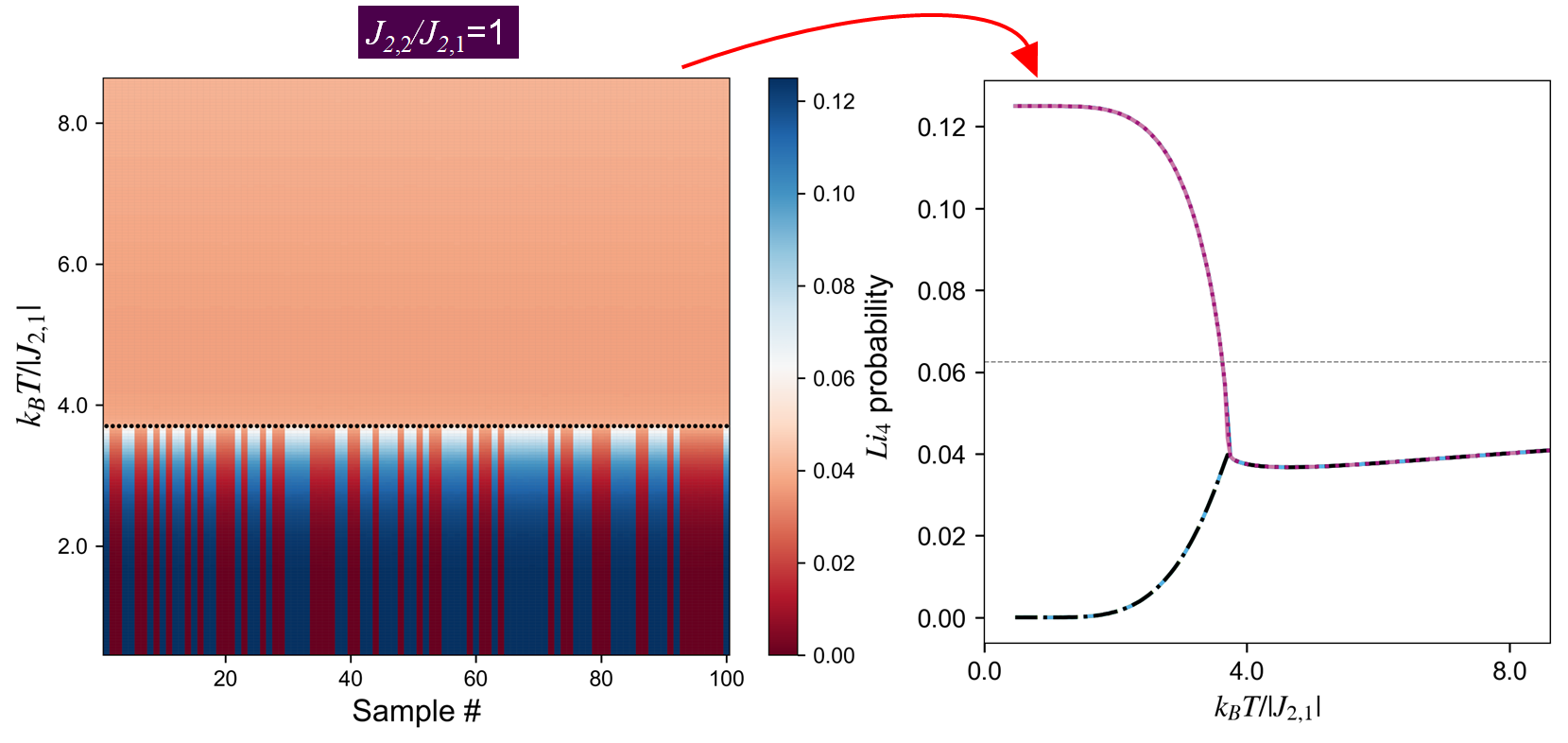}
\caption{\label{fig:SI_LvsSP} Degeneracy of Layered and Spinel-like LRO when $J_{4,1} = 0$. Simulations with $J_{2,2}$/$J_{2,1}$ $>$ 0.5 below $T_c$ have a 50\% probability of equilibrating into either Layered or Spinel-like LRO randomly, as demonstrated in this case of $J_{2,2}$/$J_{2,1} = 1$ performed 100 times. Regardless of the ground state LRO, all samples have the same high-temperature ordering parameters (including Li$_4$ probability) as shown in the plot on the right-hand side.  } 
\end{figure}

\end{document}